\documentclass[a4paper,11pt]{article}
\usepackage{fullpage,amsmath,amssymb,mathtools,natbib}
\usepackage{graphicx,bbm,url}
\usepackage[plain]{algorithm2e}
\usepackage{tikz}
\usepackage{fancyhdr,lipsum}
\usepackage{placeins,bm}
\usetikzlibrary{arrows, decorations.markings}

\def\pconv{\smash{\mathop{\longrightarrow}\limits^p}}     %Converges in prob
\def\dconv{\smash{\mathop{\longrightarrow}\limits^d}}     %Converges            %in dist.

\def\argmin{\mbox{argmin}}
\renewcommand{\bar}{\overline}

\renewcommand{\hat}{\widehat}

\newcommand{\beq}{\begin{eqnarray*}}
\newcommand{\eeq}{\end{eqnarray*}}

\def\1T{frac{1}{T}}
\def\1n{\frac{1}{n}}

\def\bitem{\medskip\begin{itemize} \itemsep=8.0pt \parskip=8.0pt}
\def\eitem{\end{itemize}}

\newtheorem{proposition}{Proposition}

\def\mymathcal{\mathbbm}
\usepackage[top=1.5in,bottom=1.in,right=1in,left=1in,headheight=75pt]{geometry}

\title{ A Likelihood-Free  Reverse Sampler of the Posterior Distribution}
\author{Jean-Jacques Forneron\thanks{Department of Economics, Columbia
University, 420 W. 118 St., New York, NY 10025. Email: jmf2209@columbia.edu} \and
Serena Ng\thanks{Department of Economics, Columbia
University, 420 W. 118 St. Room 1117, New York, NY 10025. Email Serena.Ng at
Columbia.edu.   Financial support is provided by the
 National Science Foundation, SES-0962431. \newline
 We thank Christopher Drovandi, Neil Shephard, and two anonymous referees
 for many helpful comments. The second author would like to thank  Aman Ullah  for his support and guidance. }
}
\date{October 2015}

\newcommand{\changefont}{%
    \fontsize{9}{11}\selectfont
}
\begin{document}
  \let\thetanormal\theta
  \def\theta{\bm{\thetanormal}}
  \let\psinormal\psi
  \def\psi{\bm{\psinormal}}
  \let\varthetanormal\vartheta
  \def\vartheta{\bm{\varthetanormal}}
  \let\varepsilonnormal\varepsilon
  \def\varepsilon{\bm{\varepsilonnormal}}
\maketitle

\begin{abstract}
This paper considers properties of an optimization based  sampler for targeting the posterior
distribution when the likelihood is intractable. It uses auxiliary statistics
to summarize information in the data and   does not directly evaluate the likelihood associated with the specified
parametric model.
 Our reverse sampler  approximates  the desired posterior distribution by
 first
solving a sequence  of simulated minimum distance  problems. The solutions
are then re-weighted by an importance ratio that depends on the
prior and the volume of the Jacobian matrix. By a change of variable argument,
the output are  draws from the desired
posterior distribution. Optimization
 always results in acceptable draws. Hence
when the minimum distance  problem is not too
difficult to solve, combining importance sampling with
optimization can be much faster than the method of Approximate Bayesian Computation
that by-passes optimization.
\end{abstract}



\bibliographystyle{harvard}
\newpage
\baselineskip=18.0pt
\section{Introduction}
Maximum likelihood estimation rests on the ability of a researcher to
express the
joint density of the data, or the likelihood, as a function of $K$ unknown
parameters $\theta$. Inference can be conducted using classical distributional theory   once the  mode of  the likelihood function is determined by numerical optimization.  Bayesian estimation combines the likelihood with a prior  to form the posterior distribution from which the mean and other quantities of interest can be computed. Though  the posterior  distribution may not always be tractable, it can be approximated by Monte Carlo methods provided that the likelihood is available.
When the likelihood is intractable but there exists $L\ge K$ auxiliary statistics $\hat\psi$ with model analog $\psi(\theta)$ that is analytically tractable, one can still estimate $\theta$ by  minimizing the difference between $\hat\psi$ and $\psi(\theta)$.

Increasingly,  parametric models are
so complex that neither the likelihood  nor  $\psi(\theta)$ is  tractable.
But if the model is easy to simulate,
 the mapping $\psi(\theta)$ can be approximated by simulations.
Estimators that exploit this idea  can  broadly be classified
 into two types. One is simulated
minimum distance estimator (SMD), a frequentist approach that is quite widely  used in economic analysis.
The other  is the method of Approximate
Bayesian Computation  that is popular in other disciplines.
This method, ABC for short, approximates the  posterior
distribution using auxiliary statistics $\hat\psi$ instead of the full dataset
$y$. It takes draws of $\theta$ from a prior distribution and keeps the draws
that, when used to simulate the model, produces  auxiliary statistics  that
are close to the sample estimates $\hat\psi$. Both the ABC and SMD can be
regarded as likelihood free estimators in the sense that the likelihood
that corresponds to the
structural model of interest is not directly evaluated.

While both the SMD and ABC exploit auxiliary statistics to perform likelihood
free estimation, there are important differences between them. The SMD
solves for the $\theta$ that makes $\hat\psi$ close to the average of
$\psi(\theta)$  over  many simulated
  paths of the data.  In contrast, the ABC evaluates $\psi(\theta)$ for
each draw from the prior and accepts the draw only if $\psi(\theta)$ is close
  to  $\hat\psi$. The ABC estimate is  the  average over the accepted draws,
  which is the posterior mean.
  In \citet{jjng-14}, we focused on the case of exact identification and used
  a reverse sampler (RS)   to better understand the difference between the
  two approaches. The RS approximates the posterior distribution by solving
  a sequence of SMD problems, each using only one simulated path of data.
  Using stochastic expansions as in \citet{rsu-96} and
\citet{bao-ullah:07}, we reported that in the special case when $\psi(\theta)=\theta$ (i.e the auxiliary model is the assumed model), the SMD has an unambiguous bias advantage over the ABC. But in more general settings, the ABC can, by clever choice of prior, eliminate biases that are inherent in the SMD.

In this paper, we extend the analysis to over-identified models and provide
a deeper understanding of the reverse sampler. The RS is shown to be an optimization-based
importance sampler that
 transforms the  density from draws of $\psi$
to  draws of $\theta$ so that when multiplied by the
prior and properly weighted, the draws follow the desired posterior distribution. Section 2 considers the  exactly identified case and shows that the importance ratio
is the determinant of the Jacobian matrix.
Section 3 considers the over-identified case when the dimension of $\psi(\theta)$ exceeds that of $\theta$.  Because of the need to transform  densities of different dimensions, the determinant of the Jacobian matrix is replaced by its volume.
Using analytically tractable models, we show that the RS exactly reproduces the desired posterior distribution.

The RS was initially  developed as a framework to better understand the different
approaches to likelihood free estimation. While not intended to compete
with existing implementations of ABC,
the use of optimization in RS turns out to have a property that is of independent interest.
%To evaluate the posterior distribution  conditional on the observed auxiliary statistics, ABC algorithms typically combine simulation and non-parametric methods to keep draws that are close to $\hat \psi$ as measured by $\delta$.  In theory $\delta$ should be taken as small as possible. But in practice, the Markov Chain from the Metropolis-Hastings algorithm has poor properties when  $\delta$ is small.
Creating a long sequence of  ABC draws such that the simulated statistic $\hat\psi^b$ and the
data $\hat\psi$
deviate by no more than $\delta$ can take infinite time
if $\delta$ is set to exactly zero as theory suggests.
This has generated interests within the ABC community to   control for $\delta$.
 The RS by-passes this problem because SMD estimation makes $\hat\psi^b$
 as close to $\hat\psi^b$
as machine precision permits. We elaborate on this feature in Section 4. Of course, the RS is useful only when  the
 SMD objective function is well behaved and easy to optimize, which may not
 always be the case. But allowing  optimization  to play a role in ABC can
 be useful, as independent work by \citet{meeds-welling} also found.
%Replication files are available on the authors' website.\footnote{\url{http://www.columbia.edu/sm2204/research.html}}

\subsection{Preliminaries}
 In what follows, we use a `hat' to denote estimators that correspond to
 the mode (or extremum estimators) and a `bar' for estimators that correspond to the posterior mean.
We use $(s,S)$ and $(b,B)$ to denote the (specific, total number of)
draws in frequentist  and Bayesian type analyses respectively.
 A  superscript
$s$ denotes a specific draw and  a  subscript $S$  denotes the average over
$S$ draws. These  parameters $S$ and $B$ have different roles. The SMD uses $S$ simulations to approximate the mapping $\psi
  (\theta)$, while the ABC uses $B$ simulations to approximate the posterior distribution
  of the infeasible likelihood.

%For a function $f(\theta)$,
%we use $\hat f_{\theta}(\theta_0)$ to denote $\frac{\partial}
%{\partial \theta}f(\theta)$ evaluated at $\theta_0$ and $\hat f_{\theta\theta_j}(\theta_0)$
%to denote $\frac{\partial }{\partial \theta_j} f_{\theta}
%(\theta)$ evaluated at $\theta_0$.

We assume that the data
$\mathbf y=(y_1,\ldots,y_T)^\prime
$  have finite fourth moments and can be represented by a parametric model
with probability measure $\mathcal P_{\theta}$ where $\theta \in \Theta\subset
\mymathcal R^K$,   $\theta_0$ is the true value. The likelihood $L(\theta|\mathbf y)$
 is intractable. Estimation of $\theta$ is based on $L\ge K $
auxiliary statistics $\hat\psi(\mathbf y(\theta_0)) $ which we simply denote
by $\hat\psi$ when the context is clear.
The model implies statistics $\psi(\theta)$. The classical minimum distance estimator is
\[\hat\theta_{\text{CMD}}=\argmin_{\theta} J(\hat\psi,\psi(\theta))=\bar g(\theta)^\prime \mathrm W \bar g(\theta), \quad \bar g(\theta)=\hat\psi-\psi(\theta).\]
\newpage
\paragraph{Assumption A}:
\begin{itemize}
\item[i] There exists a unique interior point $\theta_0\in \Theta$ (compact) that minimizes
the population objective function $(\psi(\theta_0)-\psi(\theta))^\prime \mathrm{W} (\psi(\theta_0)-\psi(\theta))$.
The mapping  $ \theta\rightarrow \psi(\theta)=\lim_{T\rightarrow\infty} \mathbb E[\hat\psi(\theta)]$
 is continuously differentiable and injective.
 The $L\times K$ Jacobian matrix $\psi_{\theta}(\theta)=\frac{\partial \psi(\theta)}{\partial \theta}$ has full column rank, and the rank is constant in the neighborhood of $\theta_0$.

\item[ii]  There is an estimator $\hat\psi$ such that   $\sqrt{T}(\hat\psi -\psi(\theta_0))
\dconv \mathcal N(0,\Sigma)$.

\item[iii] $\mathrm{W}$ is a $L\times L$ positive definite matrix and $\mathrm{W}\psi_{\theta}(\theta_0) $ has rank $K$.
\end{itemize}

Assumption A ensures global identification and consistent estimation of $\theta$,
see \citet{newey-mcfadden-handbook}. In
  \citet{gmr},  the mapping $\psi:\theta\rightarrow\psi(\theta)$ is referred
  to as the binding function
  while in \citet{jiang-turnbull:94},  $\psi(\theta)$ is referred to as a bridge function.
When $\psi(\theta)$ is analytically intractable, the simulated minimum distance estimator (SMD)  is
\begin{eqnarray}
\label{eq:SMD}
\hat\theta_{\text{SMD}}&=&\argmin_
{\theta} J_S(\hat\psi,\hat\psi_S(\theta))=\argmin_{\theta} \bar g_S(\theta)^{\prime} \mathrm{W} \bar g_S(\theta).
\end{eqnarray}
where $S\geq 1$ is the number of simulations,
\[   \bar g_S(\theta) =  \hat\psi -\frac{1}{S}\sum_{s=1}^S
\hat\psi^s(\mathbf y^s(\theta)).
\]
Notably, the term $\mathbb E[\hat\psi(\theta)]$  in CMD estimation is approximated
by $\frac{1}{S}\sum_{s=1}^S \hat\psi^s(\mathbf y^s(\theta))$.
The SMD was first used in \citet{smith-93}. Different SMD estimators can be obtained
by suitable choice of the moments $\bar g(\theta)$,
  including the  indirect inference estimator of  \citet{gmr}, the simulated method of moments of \citet{duffie-singleton}, and the efficient method of moments of \citet{gallant-tauchen-emm}.

The first ABC algorithm was implemented  by \citet{tbfd} and \citet{pspf:99} to study
population genetics. They   draw $\theta^b$ from the prior
distribution $\pi(\theta)$,  simulate the model  under $\theta^b$ to obtain
data $\mathbf y^b$, and  accept $\theta^b$ if the  vector of auxiliary
statistics $\psi(\theta^b)$  deviates from
 $\hat\psi$ by no more than a tuning parameter $\delta$.  If $\hat\psi$ are sufficient statistics
 and   $\delta=0$, the  procedure  produces samples from the true posterior distribution if $B\rightarrow\infty$.
 \paragraph{The Accept-Reject ABC:} For $b=1,\ldots, B$

\begin{itemize}
\item[i] Draw $\vartheta$ from $\pi(\theta)$ and $\varepsilon^b$ from an assumed
distribution $F_{\varepsilon}$
\item[ii] Generate $\mathbf y^b(\varepsilon^b,\vartheta)$ and $\hat\psi^b=\psi(\mathbf y^b)$.
\item[iii] Accept $\theta^b=\vartheta$ if $J_1^b=\left(\hat \psi^b - \hat \psi \right)^\prime \mathrm{W} \left(\hat \psi^b - \hat \psi \right) \leq \delta$.
\end{itemize}
The accept-reject method (hereafter, AR-ABC) simply keeps those draws from the prior distribution $\pi(\theta)$ that produce auxiliary statistics which are close to the observed $\hat \psi$.
As it is not easy to choose $\delta$ a priori, it is common in AR-ABC to fix
a desired quantile $q$, repeat the steps  $[B/q]$ times.
Setting $\delta$ to the $q$-th quantile of  the sequence of $J_1^b$ that will  produce exactly
 $B$ draws is analogous  to the idea of keeping $k-$nearest neighbors
 considered in \citet{gao-hong}.

 Since simulating from a non-informative prior distribution is inefficient,  the
accept-reject sampler can be replaced by one
that targets at features of the posterior distribution. There are many ways
to target the posterior distribution. We consider
the MCMC implementation of ABC  proposed in \citet{mmpt-03}  (hereafter, MCMC-ABC).

\paragraph{The MCMC-ABC:} For $b=1,\ldots, B$ with $\theta^0$ given and proposal density $q(\cdot |\theta^b)$,
\begin{enumerate}
\item[i] Generate $\vartheta \sim q(\vartheta|\theta^b)$
\item[ii] Draw errors $ \varepsilon^{b+1}$ from $F_{\varepsilon}$ and  simulate data
$\mathbf
 { y^{b+1}}(\varepsilon^{b+1},\vartheta)$. Compute
 $\hat \psi^{b+1}=\psi( \mathbf y^{b+1})$.
\item[iii] Set  $\theta^{b+1}$ to $\vartheta$  with probability $ \rho_{ABC}
(\theta^b,\vartheta) $ and  to
$\theta^{b+1}$ with probability $ 1-\rho_{ABC}(\theta^b,\vartheta)$ where
 \begin{equation}
 \label{eq:rhoABC-simple}
 \rho_{ABC}(\theta^b,\vartheta) = \min \Big( \mathbbm I_{\|\hat\psi,\hat\psi^{b+1}\|\le \delta}
 \frac{\pi(\vartheta)q(\theta^b|\vartheta)}{\pi(\theta^b)q(\vartheta|\theta^b)}
,1 \Big  )\end{equation}
\end{enumerate}
The AR and MCMC both produce an approximation to the posterior distribution of $\theta$. It is common to use the  posterior mean of the draws $\bar\theta=\frac{1}{B}\sum_{b=1}^B \theta^b$ as the ABC estimate.
The MCMC-ABC uses a proposal distribution to account for features of the data
so that it is less likely to have proposed values with low posterior probability. The tuning parameter $\delta$ affects the bias of the estimates.
Too small a $\delta$ may require making many draws which can be computationally costly.

The ABC samples from the joint distribution of $(\theta^b, \psi^b(\varepsilon^b,\theta^b))$  and then integrates out $\varepsilon^b$.    The posterior distribution is thus
\[  p (\theta^b|\hat\psi) \propto  \int p(\theta^b,\hat \psi^b(\theta^b,\varepsilon^b)|\hat\psi )
\mathbbm I_{\|\hat\psi-\psi^b)\|<\delta}d\varepsilon^b.\]
The indicator function (also the rectangular kernel) equals one if
$\|\hat\psi- \psi^{b}\|$ does not exceed
$\delta$.
%\footnote{  \citet{wilkinson-08} takes $K_\delta(\cdot)$ to be the distribution of the error, or discrepancy, between the data and the best prediction due to the model.  This (noisy) ABC simulates from the  posterior distribution that takes errors-in-variables in account.
%}
The ABC draws are dependent due to  the Markov nature of the MCMC-ABC sampler.
%The general idea of the ABC is explained in more detail in \citet{jjng-14}.

Both the SMD and ABC assume that simulations provide an accurate approximation
of $\psi(\theta)$ and that auxiliary statistics are chosen to permit identification
of $\theta$. \citet{creel-kristensen:15} suggests a cross-validation method
 for selecting the auxiliary statistics. For the same choice of $\hat\psi$,
the SMD  finds the $\theta$ that makes the average of the simulated auxiliary
statistics close to $\hat\psi$.  The ABC takes the average of  $\theta^b$,
drawn from the prior,  with the property that each $\psi^b$ is close to $\hat\psi$.
In an attempt to understand this difference,  \citet{jjng-14},
takes as starting point that each $\theta^b$ in the above ABC algorithm can be reformulated as an SMD problem with $S=1$. We consider an  algorithm  that solves the SMD
problem   many times to obtain a distribution
for $\theta^b$,  each time using  one simulated path.
 The sampler  terminates  with an evaluation of the prior probability, in
 contrast to the ABC which starts with a draw from the prior distribution.
 Hence we call our algorithm a reverse sampler (hereafter, RS).
The RS produces a sequence of  $\theta^b$  that are independent optimizers
and do not have a Markov structure.
%The RS has features of both the SMD and
%ABC and was used in \citet{jjng-14} to
%understand how different likelihood-free estimators relate to each other.

In the next two sections, we explore additional features of the RS.  As an
overview, the distribution of draws that emerge from SMD estimation with $S=1$
may not
be from the desired  posterior distribution. Hence the draws are re-weighted  to target the posterior.
 In the exactly identified case, $\hat\psi^b$ can be made exactly equal to
 $\hat\psi$ by choosing the SMD estimate as $\theta^b$. Thus the RS  is simply  an optimization based importance sampler using the determinant of Jacobian matrix as importance ratio.
In the over-identified case, the volume of the (rectangular) Jacobian matrix
is used in place of the determinant. Additional  weighting is given to those
$\hat\theta^b$ that yields $\hat\psi^b$ sufficiently close to $\hat\psi$.

\section{The Reverse Sampler: Case $K=L$}

The algorithm for the case of exact identification is as follows. For $b=1,\ldots, B$

\begin{itemize}
\item[i] Generate $\varepsilon^b $ from $F_{\varepsilon}$.
\item[ii] Find $\theta^b=\argmin_{\theta} J^b_1(\hat \psi^b(\theta,\varepsilon^b),\hat\psi)$ and
let $\hat\psi^b=\hat \psi^b(\theta^b,\varepsilon^b).$
\item[iii] Set $w(\theta^b,\varepsilon^b)=\pi(\theta^b)|\hat \psi^b_{\theta}(\theta^b,\varepsilon^b)|^{-1}$.
\item[iv] Re-weigh the $\theta^b$ by $\frac{w(\theta^b)}{\sum_{b=1}^B w(\theta^b)}.$
\end{itemize}
Like the ABC, the draws $\theta^b$ provides an estimate of the posterior distribution of $\theta$ from which an estimate of the posterior mean:
 \[ \bar\theta_{RS }=\sum_{b=1}^B \frac{w(\theta^b)}{\sum_{b=1}^B w(\theta^b)}
\theta^b \]
can be used as an estimate of $\theta$.
Each $\theta^b$ is a function of the data $\hat \psi$ and the draws $\varepsilon^b$ that minimizes $J_1^b( \psi(\theta,\varepsilon^b),\hat\psi)$. The  $K$  first-order conditions are given by
 \begin{eqnarray}
\label{eq:foc}
\mathcal F(\theta^b,\varepsilon^b,\hat\psi)= \frac{\partial \bar g_1(\theta^b,\varepsilon^b,\hat\psi)}{\partial \theta}^\prime
 \mathrm{W} \bar g_1(\theta^b,\varepsilon^b,\hat\psi)=0
 \end{eqnarray}
where $ \frac{\partial \bar g_1(\theta^b,\varepsilon^b,\hat\psi)}{\partial \theta}$  is
 the   $L\times K$ matrix of derivatives with respect to $\theta$ evaluated at the arguments.  It is assumed   that, for all $b$, this derivative matrix has full column rank $K$. For SMD estimation, $\frac{\partial \bar g_1(\theta^b,\varepsilon^b,\hat\psi)}{\partial \theta}=\hat\psi_{\theta}^b(\theta^b,\varepsilon^b,
 \hat\psi)$. This Jacobian matrix plays an important role in the RS.

The importance  density denoted  $h(\theta^b,\varepsilon^b|\hat\psi)$ is obtained
by drawing $\varepsilon^b$ from the assumed distribution $F_{\varepsilon}$  and finding $\theta^b $ such that
$J(\hat\psi^b(\theta,\varepsilon^b),\hat\psi)$ is smaller than a pre-specified
tolerance. When $K=L$, this tolerance can be made arbitrarily small so that
up to numerical precision, $\hat \psi^b(\theta^b,\varepsilon^b)=\hat\psi$.
This density $h(\theta^b,\varepsilon^b|\hat\psi)$   is related to
$p_{\hat\psi^b,\varepsilon^b}(\hat \psi^b(\theta^b,\varepsilon^b))\equiv p(\hat\psi^b,\varepsilon^b)$  by a change of variable:
\[ h(\theta^b,\varepsilon^b|\hat\psi )= p(\hat \psi^b, \varepsilon^b|\hat\psi)\cdot  |\hat \psi^b_{\theta}(\theta^b,\varepsilon^b)|.\]
% For fix $\theta$,
% the importance density of $\theta^b$ is
%\begin{align*}
%  g(\theta^b|\hat\psi) &=  \int g(\theta^b,\varepsilon^b|\hat\psi) d\varepsilon^b = \int  p(\hat \psi^b,\varepsilon^b|\hat\psi)
%    |\psi_{\theta}(\theta,\varepsilon^b)|  d\varepsilon^b.
%\end{align*}

Now $p(\theta^b,\hat\psi^b|\hat\psi)\propto p(\hat\psi|\theta^b,\hat\psi^b)p(\hat\psi^b,\varepsilon^b|\theta^b)\pi(\theta^b)$ and
$p(\hat\psi|\theta^b,\hat\psi^b)$ is constant since $\hat\psi^b=\hat\psi$.
 Hence
 \begin{eqnarray*}
  p(\theta^b|\hat\psi)  &\propto& \int \pi(\theta^b) p(\hat\psi^b,\varepsilon^b|\hat\psi )
  \mathbbm I_{\|\hat\psi-\hat\psi^b\|=0} d\varepsilon^b \\
 &=&\int \pi(\theta^b) |\hat \psi^b_{\theta}(\theta^b,\varepsilon^b,\hat\psi)|^{-1}h(\theta^b,\varepsilon^b|\hat\psi)\mathbbm I_{\|\hat\psi-\hat\psi^b\|=0}d\varepsilon^b\\
 &=& \int w(\theta^b,\varepsilon^b)h(\theta^b,\varepsilon^b|\hat\psi)d\varepsilon^b
 \end{eqnarray*}
where the weights are, assuming invertibility of the determinant:
  \begin{equation}
 \label{eq:eqw}
 w(\theta^b,\varepsilon^b)=\pi(\theta^b)|
 \hat  \psi^b_{\theta}(\theta^b,\varepsilon^b,\hat\psi)|^{-1}.
\end{equation}
  Note that in general, $\frac{w(\theta^b)}{\sum_{b=1}^B w(\theta^b)}\ne \frac1B$.

In the above, we have used the fact that   $\mathbbm I_{\|\hat\psi-\hat\psi^b\|=0}$ is 1 with probability one  when $K=L$.
The Jacobian of the transformation appears in the weights
 because the draws $\theta^b$ are related to the likelihood via a change of variable.
Hence  a crucial aspect of the RS is that it re-weighs  the draws of  $\theta^b$
from $h(\theta^b,\varepsilon)$.  Put differently, the unweighted draws will
not, in general,  follow  the target posterior distribution.

Consider a weighted sample $(\theta^b,w(\theta^b,\varepsilon))$
 with $w(\theta^b,\varepsilon^b)$ defined in (\ref{eq:eqw}).
 The following proposition shows that as $B\rightarrow \infty$,  RS  produces
 the posterior distribution associated with the infeasible likelihood,  which is also the ABC posterior distribution with $\delta=0$.

\begin{proposition}
\label{prop:prop1}
Suppose that  $\hat \psi^b: \theta\rightarrow \hat\psi^b(\theta,\varepsilon^b)$ is one-to-one and the determinant $|\frac{\partial \psi^b(\theta,\varepsilon^b,\hat\psi)}{\partial \theta}|=|\hat \psi^b_{\theta}(\theta,\varepsilon^b,\hat\psi)|$ %has  full column rank around $\theta^b$.
is bounded away from zero around $\theta^b$.
 For any measurable function $\varphi(\theta)$ such that $\mathbb{E}_{p(\theta|\hat \psi)}\left(\varphi
  \left(\theta \right)\right)= \int \varphi \left( \theta\right)p(\theta|\hat\psi)d\theta$ exists, then
    \begin{align*}
 \frac{\sum_b^B w(\theta^b,\varepsilon^b)\varphi(\theta^b)}{\sum_b^B w(\theta^b,\varepsilon^b)} \overset{a.s.}{\longrightarrow}
  \mathbb{E}_{p(\theta|\hat \psi)}\left(\varphi \left(\theta \right)\right).
\end{align*}
\end{proposition}

 Convergence to the target distribution follows from a strong law of large numbers.   Fixing the event
$\hat\psi^b=\hat\psi$ is crucial to this convergence result.
 To see why, consider first the numerator:
\begin{align*}
  \frac{1}{B}\sum_b w(\theta^b,\varepsilon^b)\varphi(\theta^b) &\overset{a.s.}{\longrightarrow} \iint \varphi \left(\theta \right) w\left(\theta,\varepsilon\right)
  p(\hat \psi^b,\varepsilon^b|\theta) |\hat \psi_{\theta} (\theta,\varepsilon,\hat\psi)|
   d\varepsilon^b d\theta \\ &= \iint \varphi \left(\theta \right) \left| \hat \psi^b_{\theta}(\theta,\varepsilon,\hat\psi)\right|^{-1} \pi(\theta) p(\hat \psi^b,\varepsilon^b|\theta)\left| \hat \psi^b_{\theta}(\theta,\varepsilon,\hat\psi)\right| d\varepsilon^b d\theta \\
  &= \iint \varphi \left(\theta \right)  \pi(\theta) p(\hat \psi^b,\varepsilon|\theta) d\varepsilon d\theta \\
  &= \iint \varphi \left(\theta \right)  \pi(\theta) p(\hat \psi,\varepsilon|\theta) d\varepsilon d\theta \\
   &= \int \varphi \left(\theta \right) \pi(\theta) L(\hat \psi|\theta) d\theta.
\end{align*}
Furthermore, the denominator converges to the integrating constant since $\frac
{1}{B}\sum_b  w(\theta^b,\varepsilon)\overset{a.s.}{\longrightarrow} \int
\pi(\theta) L(\hat\psi|\theta)d\theta$.
%\begin{align*}
%  \frac{1}{B}\sum_b  w(\theta^b,\varepsilon)&\overset{a.s.}{\longrightarrow}
%  \iint  w\left(\theta,\varepsilon\right) p(\hat \psi,\varepsilon|\theta)
%   \left| \hat \psi^b_{\theta}(\theta,\varepsilon,\hat\psi)\right| d\varepsilon d\theta \\
%  &= \iint  \left| \hat \psi^b_{\theta}(\theta,\varepsilon,\hat\psi)\right|^{-1} \pi(\theta)
%   p(\hat \psi,\varepsilon|\theta) \left|\hat  \psi^b_{\theta}(\theta,\varepsilon,\hat\psi)\right| d\varepsilon d\theta \\
%&= \iint  \pi(\theta) p(\hat \psi,\varepsilon|\theta) d\varepsilon d\theta      \\
%   &= \int  \pi(\theta) L(\hat \psi|\theta) d\theta.
%\end{align*}
 Proposition \ref{prop:prop1} implies that the weighted average of $\theta^b$ converges to the posterior mean.
Furthermore, the  posterior quantiles produced by the reverse sampler tends
to those of the infeasible  posterior distribution $p(\theta|\hat\psi)$ as $B\rightarrow\infty$.
As discussed in \citet{jjng-14}, the
ABC can be presented as an importance sampler. Hence  the accept-reject algorithm
in \citet{tbfd} and \citet{pspf:99}, as well as the Sequential Monte-Carlo approach to ABC in \citet{sisson-fan-tanaka,toni-etal}
and \citet{beaumont-cournuet-marin-robert} are all important samplers.
The RS  differs  in that it is optimization based.
It is also developed independently in \citet{meeds-welling}.

We now use  examples to illustrate how the RS works in the exactly identified case.
\paragraph{Example 1:}
Suppose we have one observation $y \sim \mathcal{N}\left( \thetanormal, 1\right)$ or $y=\thetanormal+\varepsilonnormal$,
$ \varepsilonnormal \sim
\mathcal{N}(0,1)$. The prior for $\thetanormal$ is $\thetanormal \sim \mathcal{N} \left( 0 , 1 \right)$.
By drawing, $\thetanormal^b,\varepsilonnormal^b \sim \mathcal{N}(0,1)$, we obtain $y^b=\thetanormal^b+\varepsilonnormal^b\sim \mathcal N(0,2)$. The ABC keeps $\thetanormal^b|y^b=y$. Since $(\thetanormal^b,y^b)$ are jointly normal with covariance of 1,
%$ \begin{pmatrix} 1 & 1 \\ 1 & 2\end{pmatrix} $,
we deduce that $ \thetanormal^b|y^b=y \sim \mathcal{N} ( y/2 , 1/2)$.
The exact posterior distribution for $\thetanormal$ is $\mathcal{N}(y/2,1/2)$.\\

The RS draws $\varepsilonnormal^b \sim \mathcal{N} \left( 0,1\right)$ and computes $\thetanormal^b=y-\varepsilonnormal^b$ which is $\mathcal{N}(y,1)$ conditional on $y$. The Jacobian of the transformation is $1$. Re-weighting according to the prior, we have:
\begin{align*}
  p_\text{RS}(\thetanormal|y) &\propto \phi(\thetanormal)\phi(\thetanormal-y)
  \propto \exp \begin{pmatrix} -\frac{1}{2} \left( \thetanormal^2 + (\thetanormal-y)^2 \right) \end{pmatrix}
  \propto \exp \left( -\frac{1}{2} \left( 2\thetanormal^2 - 2\thetanormal y \right) \right) \\
%  &\propto \exp \left( -\frac{2}{2} \left( \theta^2 - 2\theta y/2 \right) \right) \\
  &\propto \exp \left( -\frac{2}{2} \left( \thetanormal - y/2 \right)^2 \right). %\rednote{check}
   %&\propto \exp \left( -\frac{2}{2} \left( \rednote{\theta} - y/2 \right)^2 \right)
\end{align*}
This is the exact posterior distribution as derived above.

\paragraph{Example 2}
Suppose $y=Q(u,\thetanormal), \varepsilonnormal\sim \mathcal{U}_{[0,1]}$ and $Q$ is a quantile function that is invertible and differentiable in both arguments.\footnote{We thank Neil Shephard for suggesting the example.} For a single draw, $y$ is a sufficient statistic. The likelihood-based posterior is:
\[ p(\thetanormal|y) \propto \pi(\thetanormal)f(y|\thetanormal).\]
The RS  simulates $y^b(\thetanormal)=Q(\varepsilonnormal^b|\thetanormal)$ and sets
$ Q(\varepsilonnormal^b|\thetanormal^b) =y$. Or,  in terms of the CDF:
\[ \varepsilonnormal^b=F(y|\thetanormal^b) \]
Consider a small perturbation to $y$ holding $u^b$ fixed:
\begin{eqnarray*}
 0 &=& dy \frac{dF(y|\thetanormal^b)}{dy} +  d\thetanormal^b \frac{dF(y|\thetanormal^b)}{d\thetanormal^b}
  = dy F^\prime_y(y|\thetanormal^b) +  d\thetanormal^b F^\prime_{\thetanormal^b}(y|\thetanormal^b).
 \end{eqnarray*}
In the above, $f\equiv F_y^\prime(\cdot)$ is the density of $y$ given $\thetanormal$.  The Jacobian  is:
\[ \bigg|\frac{d\thetanormal^b}{dy}\bigg| =  \bigg|\frac{F_y^\prime(y|\thetanormal^b)}{F^\prime_{\thetanormal^b}(y|\thetanormal^b)}\bigg|=
\bigg|\frac{f(y|\thetanormal^b)}{F_{\thetanormal^b}^\prime(y|\thetanormal^b)}\bigg| .\]
To find the distribution of $\thetanormal^b$ conditional on $y$, assume $F(y,.)$ is increasing in $\thetanormal$:
\begin{align*}
  \mathbb{P} \left( \thetanormal^b \leq t|y \right)  &= \mathbb{P}\bigg( F(y|\thetanormal^b) \leq F(y|t)|y \bigg) \\
  &= \mathbb{P}\left( \varepsilonnormal^b \leq F(y|t)|y \right)\\
  &= F(y|t).
\end{align*}
By construction, $f(\thetanormal|y)=F^\prime_{\thetanormal}(y|\thetanormal)$.\footnote{ If $F(y,\cdot)$ is decreasing in $\thetanormal$, we have $\mathbb P(\thetanormal^b\le t|y)= 1-F(y,t)$.} Putting things together,\footnote{An alternative derivation is to note that
$t=  \mathbb{P} \left( u \leq t | y \right)  = \mathbb{P}\left( u=F(y,\thetanormal^b) \leq t | y \right)
  = \mathbb{P}\left( \thetanormal^b \leq F^{-1}(y,t) = t^\prime |y \right).$
Hence $f(\thetanormal^b|y)=\frac{dt}{dt^\prime}=\frac{1}{(F^{-1})^\prime_{\thetanormal}(y,t)}=F^\prime_2(y,t)$ as above.}

\[ p_\text{RS}(\thetanormal|y) \propto \pi(\thetanormal)|F^\prime_{\thetanormal}(y|\thetanormal)|\bigg|
 \frac{f(y|\thetanormal)}{F^\prime_{\thetanormal}(y|\thetanormal)} \bigg|= \pi(\thetanormal)f(y|\thetanormal) \propto p(\thetanormal|y).\]

\paragraph{Example 3: Normal Mean and Variance}
We now consider an example in which the estimators can be derived analytically, and given in \citet{jjng-14}.  We assume
$ y_t=\varepsilon_t\sim N(
 m,\sigma^2) $.
 The parameters of the model are
 $\theta=(m,\sigma^2)^\prime.$
 We consider the auxiliary statistics:
$ \hat\psi(\mathbf y)^\prime
 =\begin{pmatrix} \bar y &  \hat\sigma^2
 \end{pmatrix}.
$
The parameters are exactly identified.

The MLE of $\theta$ is
\[\hat m=\frac{1}{T}\sum_{t=1}^T y_t, \quad\quad \hat\sigma^2= \frac{1}{T}\sum_{t=1}^T (y_t-\bar y)^2.\]
We consider the  prior $\pi(m,\sigma^2)= (\sigma^2)^{-\alpha} \mathbbm I_{\sigma^2>0} $, $\alpha>0$ so that
the log posterior distribution is
\[ \log p(\theta|\hat m,\hat\sigma^2 )\propto \frac{-T}{2}\log (2\pi) \sigma^2 -\alpha \log \sigma^2
- \frac{1}{2\sigma^2}\sum_{t=1}^T (y_t-m)^2.\]
Since $\hat\psi(\mathbf y)$ are sufficient statistics, the RS coincides with the likelihood-based Bayesian estimator, denoted $B$ below. This is also the infeasible ABC estimator.  We focus discussion on estimators for $\sigma^2$ which have more interesting properties.
 Under a uniform prior, we obtain
\begin{eqnarray*}
\bar\sigma^2_{B}&=&\hat\sigma^2 \frac{T}{T-5}\\
      \hat\sigma^2_{\text{SMD}} &=&  \frac{\hat
  \sigma^2}{\frac{1}{ST}\sum_{s=1}^S\sum_{t=1}^T (\varepsilon_t^s- \bar \varepsilon^s)^2}\\
  \hat\sigma^2_{RS} &=&\sum_{b=1}^B \frac{\frac{\hat{\sigma}^2}{[\sum_{t=1}^T(\varepsilon_t^b-\bar{\varepsilon}^b)^2/T]^2}}{\sum_{k=1}^B \frac{1}{\sum_{t=1}^T(\varepsilon_t^k-\bar{\varepsilon}^k)^2/T}}
\end{eqnarray*}
In this example, the RS is also the ABC estimator with $\delta=0$.
It is straightforward to show that the bias reducing prior is $\alpha=1$ and coincides with the SMD.
Table \ref{tbl:simple} shows that the estimators are asymptotically equivalent but can differ for fixed $T$.

\begin{figure}[ht]

  \centering
  \caption{Normally Distributed data}
  \label{fig:normal}
  \includegraphics[width=6in,height=3.5in]{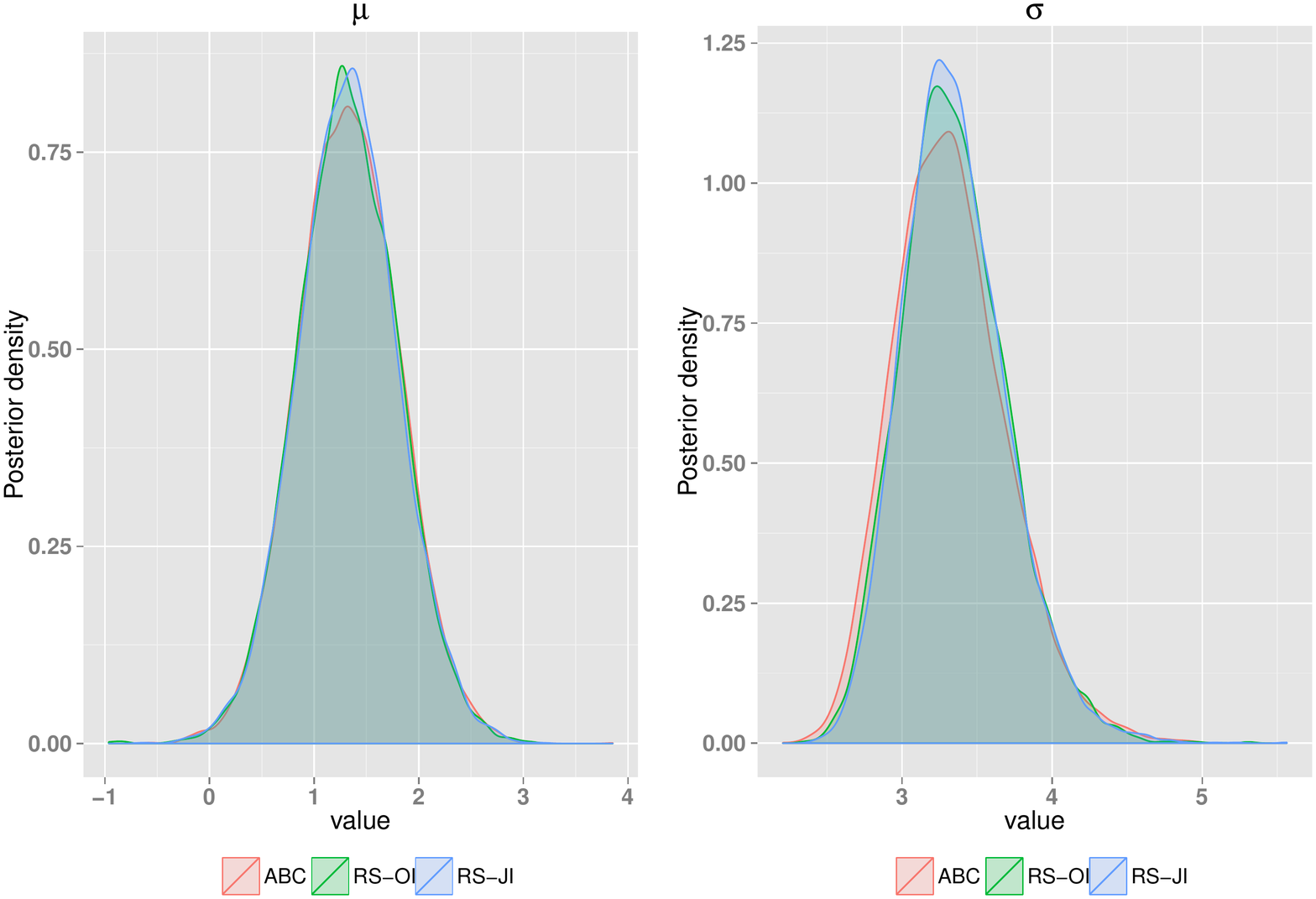}
%\end{figure}

%\begin{figure}[ht]
\caption{The Importance Weights in RS}
\label{fig:fig1}
\begin{center}
\includegraphics[width=6in,height=3.5in]{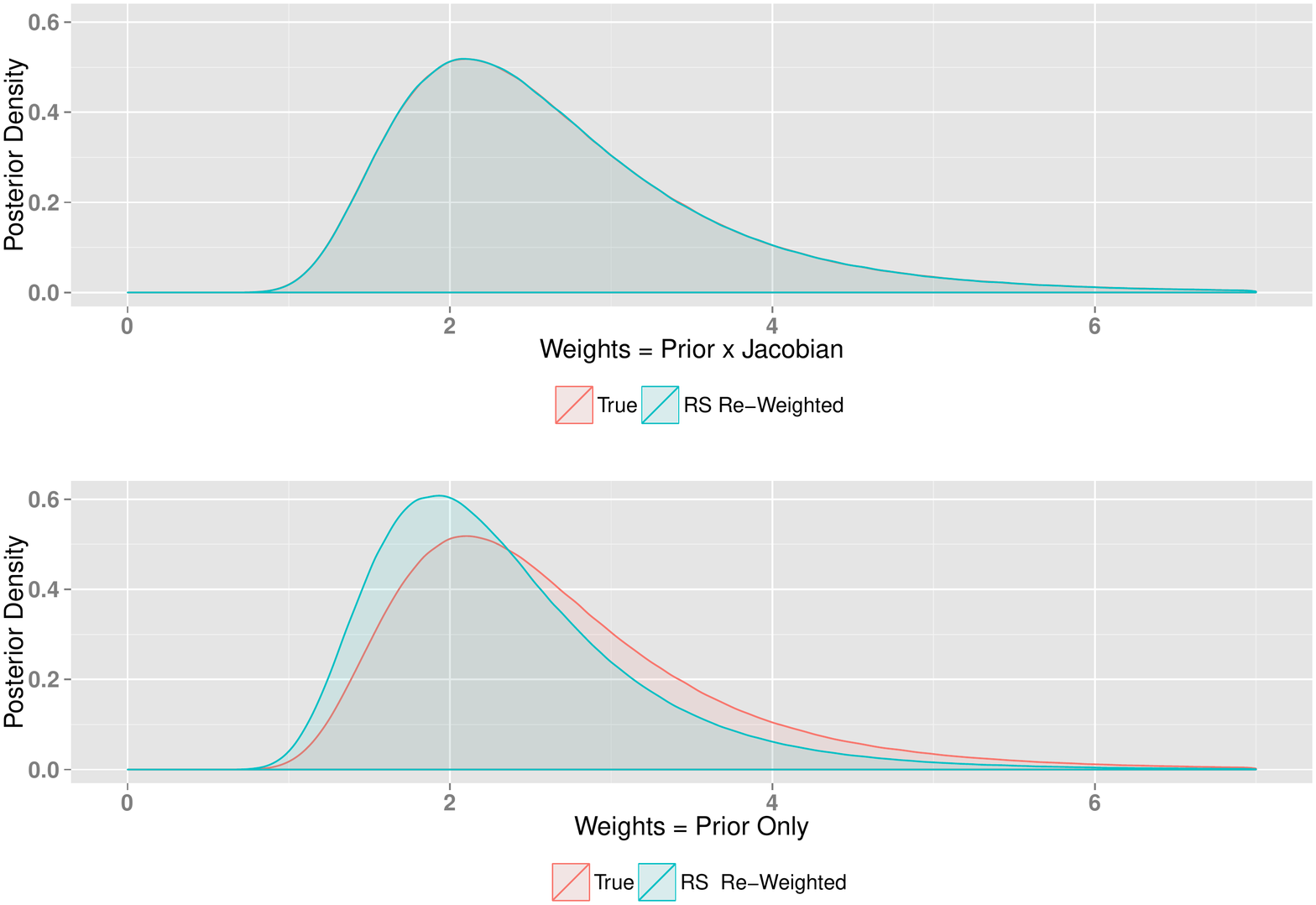}
\end{center}
\end{figure}

\begin{table}[ht]
\label{tbl:simple}
\caption{Properties of the Estimators}
\begin{center}
 \begin{tabular}{ll|l|l|l|ll}
 Estimator & Prior & $\mathbf E[\hat\theta]$  & Bias & Variance & MSE\\ \hline
 $\hat\theta_{ML}$ &-&
   $\sigma^2 \frac  {T-1}{T}$  & $-\frac{\sigma^2}{T}$ & $2\sigma^4\frac{T-1}{T
   ^2} $ & $ 2\sigma^4 \frac{2T-1}{2T^2}$
  \\
%  MD &     $\sigma^2 $ & 0 & $2\sigma^4 \frac{1}{T-1}$ & $2\sigma^4 \frac{1}{T-1}$   \\
  $\bar\theta_{B}$ & 1 & $\sigma^2\frac{T-1}{T-5} $ & $\frac{2\sigma^2}{T-5} $ & $2\sigma^4\frac
  {T-1}{(T-5)^2}$ & $2\sigma^4 \frac{T+1}{(T-5)^2}$\\
$\bar\theta_{RS}$ & 1 & $\sigma^2\frac{T-1}{T-5} $ & $\frac{2\sigma^2}{T-5} $ & $2\sigma^4\frac
  {T-1}{(T-5)^2}$ & $2\sigma^4 \frac{T+1}{(T-5)^2}$\\
    $\hat\theta_{\text{SMD}}$ & - &  $\sigma^2\frac{S(T-1)}{S(T-1)-2}$ & $\frac{2\sigma^2}{S(T-1)-2}$ &
$  2\sigma^4 \kappa_1\frac{1}{T-1}$ & $2\sigma^4 \frac{\kappa_1 }{T-1} + \frac{4\sigma^4}{(S(T-1)-2))^2}$ \\
  \hline
  \end{tabular}

\end{center}
\noindent where  $\kappa_1(S,T)= \frac{(S(T-1))^2(T-1+S(T-1)-2)}{(S(T-1)-2)^2(S(T-1)-4)}>
1$, $\kappa_1$ tends to
one as $S$  tend to infinity.
\end{table}
To highlight the role of the Jacobian matrix in the RS,   the top panel of Figure \ref{fig:fig1} plots the exact posterior
distribution and the one obtained from the reverse sampler. They are
indistinguishable. The bottom panel shows  an incorrectly constructed reverse
sampler that does not apply the Jacobian transformation.  Notably, the two
distributions are not the same. Re-weighting by the Jacobian matrix is crucial to targeting the desired posterior distribution.

Figure \ref{fig:normal} presents the likelihood based posterior distribution, along with the likelihood free ones produced by ABC and the RS-JI (just identified) for one draw of the data.
The ABC results are based on the accept-reject algorithm.
The numerical results corroborate with
 the analytical ones:  all the posterior distributions  are very similar.
The RS-JI posterior distribution  is very close to the exact posterior distribution.
 Figure \ref{fig:normal} also presents results for the over-identified case (denoted RS-OI) using  two additional auxiliary statistics:  $\hat \psi = (\bar y, \hat \sigma_y^2, \hat \mu_3/\hat \sigma_y^2, \hat \mu_4/\hat \sigma_y^4)$ where $\mu_k = \mathbb{E}(y^k)$. The weight matrix is $\text{diag}(1,1,1/2,1/2)$. The posterior distribution is very close to  RS-JI obtained for exact identification.
We now explain how the posterior distribution for the  over-identified case is obtained.

\section{The RS: Case  $L\ge K$:}
The idea behind the RS is the same when we go from the case of exact to overidentification.
The precise implementation  is as follows.
Let $\mathbb K_\delta(\hat\psi,\hat\psi^b)$ be a kernel function and $\delta$
be a tolerance level such that
$\mathbb K_0(\hat\psi,\psi^b)=\mathbbm I_{\|\hat\psi-\hat\psi^b\|=0}$.

For $b=1,\ldots, B$

\begin{itemize}
\item[i] Generate $\varepsilon^b $ from $F_{\varepsilon}$.
\item[ii] Find $\theta^b=\argmin_{\theta} J^b_1(\hat\psi^b,\hat\psi)$ where $\hat\psi^b= \hat \psi(\theta,\varepsilon^b)$;
\item[iii] Set $w(\theta^b,\varepsilon^b)=\pi(\theta^b)\text{vol}\left(\hat \psi_{\theta}^b(\theta^b,\varepsilon^b,\hat\psi)\right)^{-1}\mathbb K_\delta(J^b_1(\hat\psi^b,\hat\psi))$ where:
$\text{vol}(\hat \psi^b_{\theta})=\sqrt{\left|\hat \psi_{\theta}^{b \prime}  \hat
\psi_{\theta}^b\right| }.$
\item[iv] Re-weigh $\theta^b$ by $\frac{w(\theta^b)}{\sum_{b=1}^B w(\theta^b)} $.

\end{itemize}
We now proceed to explain the two changes:-  the use of volume in place of determinant in the importance ratio, and  the need for
 $L-K$ dimensional kernel  smoothing.

The usual change of variable formula evaluates the absolute value of the determinant of the Jacobian
matrix when the matrix is square. The determinant then gives the infinitesimal dilatation of the volume element in passing from one set of variables to another.
The main issue in the case of overidentification is that the determinant of
a rectangular Jacobian matrix
is not well defined. However, as shown in \citet{benisrael-99},  the determinant   can be replaced by the volume  when  transforming from sets of a higher dimension to a lower one.\footnote{From \citet{benisrael-01},
 $\int_V f(v)d v=\int_U f (\phi(u))\text{vol}\bigg( \phi_u(u)\bigg) du$ for
 a real valued function $f$ integrable on $V$.  See also \url{http://www.encyclopediaofmath.org/index.php/Jacobian}.}
 For a $L\times K$ matrix $A$, its volume, denoted $\text{vol}(A),$ is the
 product of the (non-zero) singular values of $A$:
 \begin{eqnarray*}
  \text{vol}( A)= \begin{cases} \sqrt{| A^\prime A}|  \quad & L \geq K, \;\;\text{rank}(A)=K\\
  \sqrt{|AA^\prime|} \quad & L \leq K,\;\; \text{rank}(A)=L.
  \end{cases}
  \end{eqnarray*}
Furthermore, if $A=BC$, $\text{vol}(A)=\text{vol}(B)\text {vol}(C).$
%The two quantities are identical when the transformation involves sets of the same size.

To verify that our target distribution is unaffected by whether we calculate the
 volume or the determinant of the Jacobian matrix  when $K=L$,
 observe that
\begin{equation}
\label{eq:identity}
\hat \psi^b_{\theta}(\theta^b(\hat\psi),\varepsilon^b)=\frac{\partial \hat  \psi^b(\theta^b,\varepsilon^b,\hat\psi)}{\partial \hat\psi} \frac{\partial\hat\psi}{\partial\theta^b}.
\end{equation}
The
$K$  first order conditions defined by (\ref{eq:foc}) become:
 \begin{equation}
 \label{eq:foc1}
\mathcal F(\theta^b,\varepsilon^b,\hat\psi)= \hat \psi^b_{\theta}(\theta^b,\varepsilon^b,\hat\psi)^\prime \mathrm{W}\bigg(\hat\psi-\hat \psi^b(\theta^b,\varepsilon^b)\bigg)=0.
 \end{equation}
 Since $L=K$,  $W$  can be set to an  identity matrix  $ I_K$.  Furthermore,
 $\psi(\theta^b,\varepsilon)=\hat\psi$  since $J_1^b(\theta^b)=0$ under exact
 identification. As
$\frac{\partial \theta}{\partial \hat\psi}$ is a square matrix when $K=L$,
we can  directly   use the fact that $\mathcal F_{\theta}(\theta^b,\varepsilon^b,\hat\psi) d\theta + \mathcal F_{\psi}(\theta^b,\varepsilon^b,\hat\psi) d\hat\psi=0$
to obtain the required determinant:
\begin{equation}
|\hat \psi^b_{\theta}(\theta^b,\varepsilon^b,\hat\psi)|^{-1}=\mathrm{I}_K\cdot |\frac{\partial \theta}{\partial \hat\psi}|=|
 -\mathcal F_{\theta}(\theta^b,\varepsilon^b,\hat\psi) ^{-1} \mathcal F_{\hat\psi}(\theta^b,\varepsilon^b,\hat\psi)|.
 \label{eq:det}
 \end{equation}
Now to use the volume result, put
  $\mathrm{A}=\mathrm{I}_K$,  $\mathrm{B}=\frac{\partial \theta}{\partial \hat\psi}$ and $\mathrm{C}=\frac{\partial \hat\psi}{\partial \theta}$. But $\mathrm{A}$ is just a $K$-dimensional identity matrix. Hence
$ \text{vol}(\mathrm{I}_K)=
 \text{vol}\bigg(\frac{\partial \theta}{\partial\hat\psi}\bigg)
 \text{vol}\bigg(\frac{\partial \hat\psi}{\partial\hat\theta}\bigg)
$ which evaluates to
 \begin{eqnarray*}
\text{vol}\bigg(\frac{\partial \hat\psi}{\partial \theta}\bigg)^{-1}=
\text{vol}\bigg(\frac{\partial \theta}{\partial
 \hat \psi}\bigg), \quad\quad\text{or} \quad\quad\bigg|\frac{\partial \hat\psi}{\partial \theta}\bigg|^{-1}=\bigg|\frac{\partial \theta}{\partial
 \hat \psi}\bigg|
 \end{eqnarray*}
which is precisely $|\hat \psi^b_{\theta}(\theta,\varepsilon)|^{-1}$ as given in
(\ref{eq:det})\footnote{Using the implicit function theorem to compute the gradient gives the same result. Since $\hat \psi^b=\hat \psi$ we have: $\mathcal F_{\theta} = -\hat \psi^b_{\theta}(\theta^b,\varepsilon^b,\hat\psi)^\prime \mathrm{W} \hat \psi^b_{\theta}(\theta^b,\varepsilon^b,\hat\psi) + \sum_j \hat \psi^b_{\theta,\theta_j}(\theta^b,\varepsilon^b)W\left(\hat \psi - \hat \psi^b(\theta^b,\varepsilon^b,\hat\psi) \right)=-\hat \psi^b_{\theta}(\theta^b,\varepsilon^b,\hat\psi)^\prime W \hat \psi^b_{\theta}(\theta^b,\varepsilon^b,\hat\psi) $.
Then $\text{vol}(\mathcal F_{\theta}^{-1}\mathcal F_{\hat \psi})=\text{vol}(\mathcal F_{\theta}^{-1})\text{vol}(\mathcal F_{\hat \psi})=\text{vol}(\hat \psi^b_{\theta}(\theta^b,\varepsilon^b,\hat\psi))^{-1}| W |^{-1} \text{vol}(\hat \psi^b_{\theta}(\theta^b,\varepsilon^b,\hat\psi))^{-1}\text{vol}(\hat \psi^b_{\theta}(\theta^b,\varepsilon^b,\hat\psi))^{-1}| W|=\text{vol}(\hat \psi^b_{\theta}(\theta^b,\varepsilon^b,\hat\psi))^{-1}$.
Hence the weights are the same when we only consider the draws where $J_1^b=0$ which are the draws we are interested in.}.
Hence in the exactly identified case, there is no difference whether one evaluates
the determinant or the volume of the Jacobian matrix.

Next, we turn to the role of the kernel function $\mathbb K_\delta
(\hat\psi,\hat\psi^b)$.
The joint density $h(\theta^b,\varepsilon^b)$ is related to $p_{\hat\psi^b,\varepsilon^b}(\hat \psi(\theta^b,\varepsilon^b))=p(\hat\psi^b,\varepsilon^b) $ through a change a variable  now expressed in terms of volume:
\[ h(\theta,\varepsilon^b|\hat\psi)=p(\hat \psi^b,\varepsilon^b|\hat\psi) \cdot \text{vol}
\left(\hat \psi_{\theta}^b(\theta^b,\varepsilon^b,\hat\psi)\right)
\]
When $L\ge K$, the objective function $\|\hat\psi-\hat\psi^b\|_{\mathrm{W}}=J_1^b\geq 0$  measures  the extent to which  $\hat\psi$ deviates from $\hat\psi^b$ when the objective function at its minimum. Consider the thought experiment that $J_1^b=0$ with probability 1, such as enabled by a particular draw of $\varepsilon^b$. Then the arguments above for $K=L$ would have applied. We would still have
$ p(\theta^b|\hat\psi)=\int \pi(\theta^b)p(\hat\psi^b,\varepsilon^b|\hat\psi)
\mathbbm I_{\|\hat\psi-\hat\psi^b\|=0} d\varepsilon^b
=\int w(\theta^b,\varepsilon^b)h(\theta^b,\varepsilon^b|\hat\psi)d\varepsilon^b$,
except that  the weights are now defined in terms of volume.
%$w(\theta^b,\varepsilon^b)=\pi(\theta^b)
%\text{vol}\bigg(\psi_{\theta}^b(\theta^b,\varepsilon^b,\hat\psi)\bigg)^{-1}
%\mathbbm I_{\|\hat\psi-\hat\psi^b\|=0}.$
Proposition 1 would then extend to the case with $L\ge K$.
%It would then follow also that
% $\frac{1}{B}\sum_b w(\theta^b,\varepsilon^b)\varphi(\theta^b)\asconv \int \varphi(\theta)\pi(\theta) L(\hat\psi|\theta)d\theta$ as in the $K=L$ case.

 But in general $J_1^b\ne 0$ almost surely. Nonetheless, we can use only  those draws that yield $J_1^b(\theta^b)$ that are sufficiently close to zero.
 The more draws we make, the tighter this criterion can be. Suppose there is  a symmetric kernel $\mathbb K_{\delta}(\cdot)$ satisfying
 %$\mathbb K_{\delta(B)}(\cdot)$
%such that $\mathbb K_\delta(B)(\hat\psi,\hat\psi^b)\rightarrow \mathbbm I_{\|\hat\psi-\hat\psi^b\|=0}$ as the
 %bandwidth
 %$\delta(B)\rightarrow 0 $  as $B\rightarrow\infty$, also assume that $\int \mathbb K(x)dx=1$, $\int x\mathbb K(x)dx=0$ and $\int x^2\mathbb K(x)dx<\infty$. The following is adapted from
 conditions in
 \citet[p.96]{pagan-ullah} for  consistent estimation of conditional moments  non-parametrically. Analogous to Proposition \ref{prop:prop1}, the volume $\text{vol}\big(\hat \psi_{\theta}^b(\theta^b,\varepsilon^b,\hat\psi)\big)$ is assumed to be bounded away from zero.
% \paragraph{Assumption B:}
% \begin{enumerate}
%   \item[i.]  $\delta(B)B \to \infty$ as $B \to \infty$.
%   \item[ii.] The density of the draws $\theta^b$ is twice continuously differentiable and bounded away from zero and infinity in a neighborhood of $\hat \psi^b=\hat \psi$.
%   \item[iii.] The function of interest $\hat \psi \rightarrow \mathbb E_{\pi} \left( \varphi(\theta)|\hat \psi\right)$ is twice continuously differentiable in $\hat \psi$ in a neighborhood of $\hat \psi.$
% \end{enumerate}
Then as the number of draws $B\rightarrow\infty$, the bandwidth $\delta(B)\rightarrow 0$ and $B\delta(B)\rightarrow\infty$ with
\begin{eqnarray}
\label{eq:w-lgtl} w_{\delta(B)}(\theta^b,\hat \varepsilon^b)&=&\pi(\theta^b)
\text{vol}\bigg(\hat \psi_{\theta}^b(\theta^b,\varepsilon^b,\hat\psi)\bigg)^{-1}
\mathbb K_{\delta(B)}(\hat\psi,\hat\psi^b),
\end{eqnarray}
a result analogous to Proposition 1 can be obtained:
\begin{eqnarray*}
&&\frac{1}{B}\sum_b w_{\delta(B)}(\theta^b,\varepsilon^b) \varphi(\theta^b)\pconv
\iint\varphi(\theta)w_0(\theta,\varepsilon) \text{vol}\bigg(\hat \psi^b_{\theta}(\theta,\varepsilon^b;\hat\psi)\bigg)
 p(\hat\psi,\varepsilon^b|\theta)d\theta d\varepsilon^b\\
 &&=
 \iint\varphi(\theta)\pi(\theta)
 \mathbbm 1_{\|\hat\psi-\hat\psi^b\|=0} \text{vol}\bigg(\hat \psi^b_{\theta}(\theta,\varepsilon^b;\hat\psi)\bigg)^{-1}
 p(\hat\psi^b,\varepsilon^b|\theta) \text{vol}\bigg(\hat \psi^b_{\theta}(\theta,\varepsilon^b;\hat\psi)\bigg)d\theta d\varepsilon^b\nonumber\\
 &&=
 \iint\varphi(\theta)\pi(\theta)
 1_{\|\hat\psi-\hat\psi^b\|=0}
 p(\hat\psi,\varepsilon^b|\theta)d\theta d\varepsilon^b\nonumber\\
 &&= \int\varphi(\theta)\pi(\theta)L(\hat\psi|\theta)d\theta.\nonumber
\end{eqnarray*}
Similarly, the integrating constant is consistent as
$\frac{1}{B}\sum_b w_{\delta(B)}(\theta^b,\varepsilon^b) \pconv\int \pi(\theta)L
(\hat\psi|\theta)d\theta.$
Hence, the RS sampler still recovers the posterior distribution with the infeasible likelihood.
Note that the kernel function was introduced for developing a result analogous to Proposition 1, but no kernel smoothing is required in practical implementation. What is needed for the RS in the over-identified case is $B$ draws with   sufficiently small $J_1(\theta^b)$. Hence,  we can borrow the idea used in the AR-ABC. Specifically, we fix a quantile $q$, repeat $[B/q]$ times until the desired number of draws is obtained.
Discarding some draws seems necessary in many ABC implementations.

In summary, there are  two changes in implementation of the RS
in the over-identified case: the volume and the kernel function.  Kernel smoothing
has no role in the RS when $K=L$.
It is interesting to note that while the ABC and RS
both rely on the kernel $\mathbb K_\delta$ to keep draws close to $\hat \psi^b$  in the over-identified case,
the  non-parametric rate at which the sum converges to the integral are different.
The RS uses the first order conditions $\hat \psi_{\theta}^b(\theta^b,\varepsilon^b)^\prime
\mathrm{W} \left( \hat \psi^b(\theta^b,\varepsilon^b) - \hat \psi \right)=0$ to indicate
which  $K$ combinations of $\hat \psi^b(\theta^b,\varepsilon^b) - \hat \psi$
are set  to zero, rendering the dimension of the smoothing problem
 $L-K$.
To see this, note first that  each draw $\theta^b$ from the
RS is consistent for $\theta_0$ and asymptotically normal as shown in \citet
{jjng-14}. In consequence, the first order condition (FOC) can be re-written
as: $\left( \frac{d\psi(\theta)}{d\theta}\big|_{\theta=\theta_0}+O_p(\frac
{1}{\sqrt{T}}) \right)^\prime \mathrm{W} \left( \hat \psi^b(\theta^b,\varepsilon^b) - \hat \psi \right)=0$, or
\[ \frac{d\psi(\theta)}{d\theta}\big|_{\theta=\theta_0}^\prime \mathrm{W} \left( \hat \psi^b(\theta^b,\varepsilon^b) - \hat \psi \right) = o_p(\frac{1}{\sqrt{T}}). \]
Since $\frac{d\psi(\theta)}{d\theta}\big|_{\theta=\theta_0}^\prime \mathrm{W}$ is full
rank, there exists a subspace of dimension $K$ such that $\hat \psi^b(\theta^b,\varepsilon^b)
- \hat \psi$ is zero asymptotically.  Hence the kernel smoothing
problem is effectively $L-K$ dimensional.
The ABC does not use  the FOC. Even in the exactly identified case, the kernel
smoothing is a $L=K$ dimensional problem. In general, the convergence rate of
the ABC is $L\geq K$,
the dimension of $\hat\psi$.

The following two examples  illustrate the properties of the ABC and RS posterior
distributions. The first example  uses sufficient statistics and the second
example does not. Both the ABC and RS  achieve the desired number of draws
by setting the quantile, as discussed in Section 2.

\paragraph{Example 4: Exponential Distribution}
Let $y_1,\dots,y_T \sim \mathcal{E}(\thetanormal), T=5, \thetanormal_0=1/2$. Now  $\hat \psinormal=\bar y$  is a sufficient statistic for $y_1,\dots,y_T$. For a flat prior $\pi(\thetanormal) \propto 1_{\thetanormal \geq 0}$  we have:
\begin{align*}
  p(\thetanormal|\bar y ) \propto p(\thetanormal|y_1,\dots,y_T) = \thetanormal^T\exp(-\thetanormal^T\bar y) \sim \Gamma(T+1,T\bar y)
\end{align*}
In the just identified case, we  let $u^b_t \sim \mathcal{U}_{[0,1]}$ and $y^b_t = -\log(1-u^b_t)/\thetanormal^b$.
This gives:
\[\hat \psinormal^b = \frac{1}{T}\sum_{t=1}^T y_t^b= -\frac{1}{T}\sum_{t=1}^T \frac{\log(1-u^b_t)}{\thetanormal^b}.\]
Since $\bar y^b=\bar y$, the Jacobian matrix is:
\[\hat \psinormal_b(\thetanormal^b)= \frac{d  \hat \psinormal^b(\thetanormal)}{d \thetanormal}\bigg|_{\thetanormal^b}
= \frac{1}{T}\sum_{t=1}^T \frac{\log(1-u^b_t)}{[\thetanormal^b]^2}= -\frac{ \bar y}{\thetanormal^b}.
\] Hence for a given $T$, the weights are:
$w(\thetanormal^b,u^b) \propto \mathbbm I_{\theta^b\ge 0}\frac{\thetanormal^b}{
\bar y^b}= \frac{\thetanormal^b}{\bar y}$. We verified that the numerical results agree with this analytical result.

In the over identified case,
we consider two moments:
\[\hat \psi^b = \begin{pmatrix} \bar y^b\\  \hat \sigma_y^{b,2}\end{pmatrix}
 = \begin{pmatrix} \frac{1}{T}\sum_{t=1}^T y_t^b\\
 \frac{1}{T}\sum_{t=1}^T (y_t^b)^2- (\frac{1}{T}\sum_{t=1}^T y_t^b)^2\end{pmatrix}. \]
Since $\frac{d y_t^b}{d\thetanormal}= \frac{\log(1-u_t^b)}{(\thetanormal^b)^2}= -\frac{ y_t^b}{\thetanormal^b}$.
If  $\delta=0$, the Jacobian matrix  is
 \[   \hat \psi^b_{\thetanormal} =
- \begin{pmatrix}
 \frac{1}{T}\sum_{t=1}^T  \frac{y_t^b}{\thetanormal}\\
  \frac{2}{\thetanormal^b}\frac{1}{T}\sum_{t=1}^T (y_t^b)^2 - \frac{2}{\thetanormal^b}\bigg[\frac{1}{T}
  \sum_{t=1}^T y_t^b\bigg]^2
\end{pmatrix} =
- \begin{pmatrix} \frac{\bar y}{\thetanormal^b}  \\
 \frac{2(\hat\sigma_y)^2}{\thetanormal^b}
 \end{pmatrix}.
  \]
The volume to be computed is
$\text{vol}(\hat \psi_{\thetanormal}^b)= \sqrt{ |\hat \psi^{b \prime}_{\thetanormal}  \hat \psi^b_{\thetanormal}| }
$, as stated in the algorithm.
Even if $\mathrm W=\mathrm{I}$, the volume is   the determinant of $\hat \psi_
{\thetanormal}^b$ in the exactly identified case, plus a term relating to the variance of $y^b$. We computed $\hat \psi^b_{\thetanormal}$ for draws with $J_1^b\approx 0$ using numerical differentiation\footnote{In practice, since the mapping $\thetanormal \rightarrow \hat \psi^b(\thetanormal)$ is not known analytically, the derivatives are approximated using finite differences: $\partial_{\thetanormal_j} \hat \psi^b(\thetanormal) \simeq \frac{\hat \psi^b(\thetanormal+e_j\varepsilonnormal)-\hat \psi^b(\thetanormal-e_j\varepsilonnormal)}{2\varepsilonnormal}$ for $\varepsilonnormal \simeq 0$.} and verified that the values are very close to the ones computed analytically for this example.

\begin{figure}[ht]

  \centering
  \caption{Exponential Distribution}
  \label{fig:exponential}
  \includegraphics[width=6in, height=3.50in]{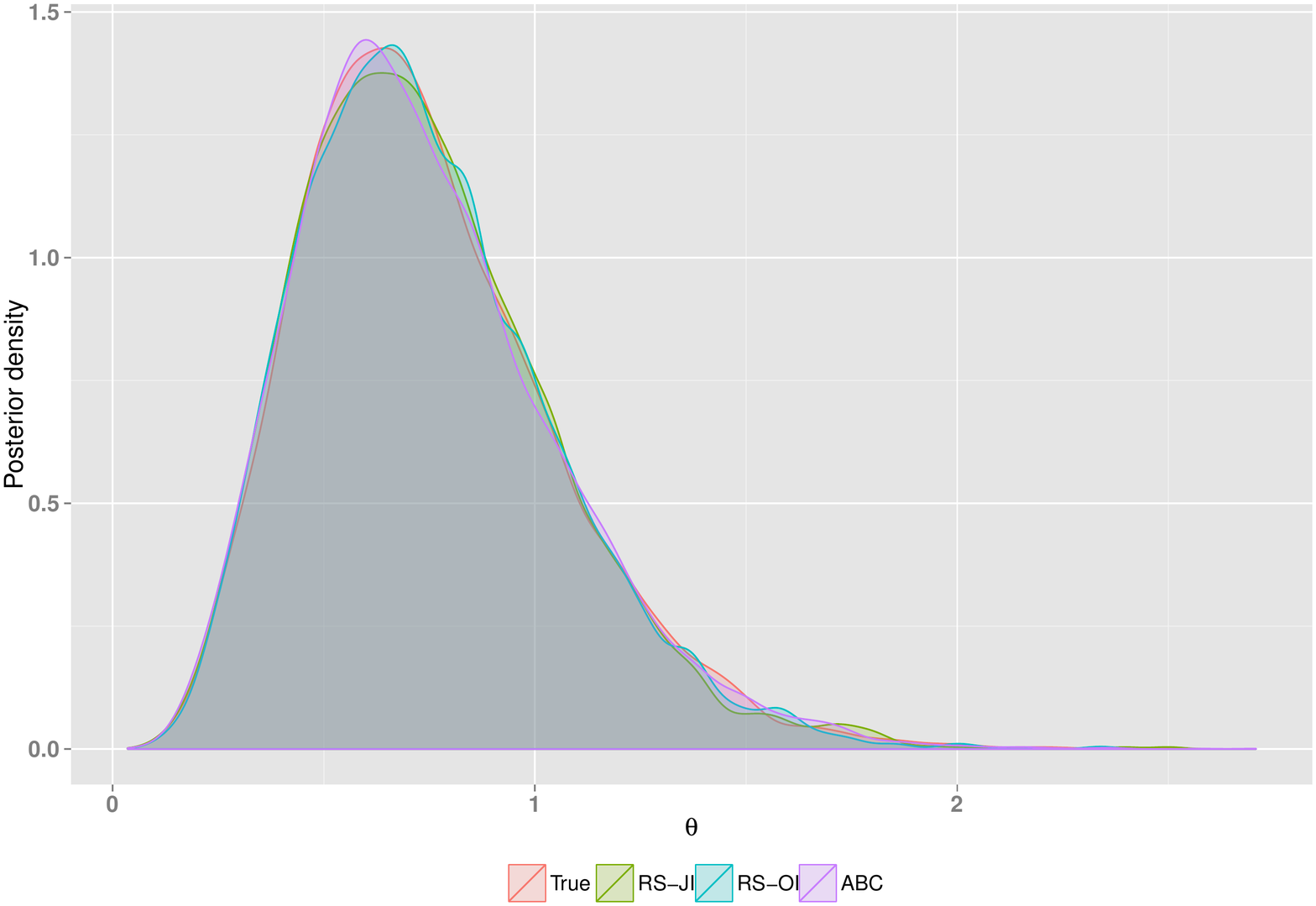}\\

  \centering
  \caption{ARMA Model}
  \label{fig:arma}
\includegraphics[width=6.0in,height=3.5in]{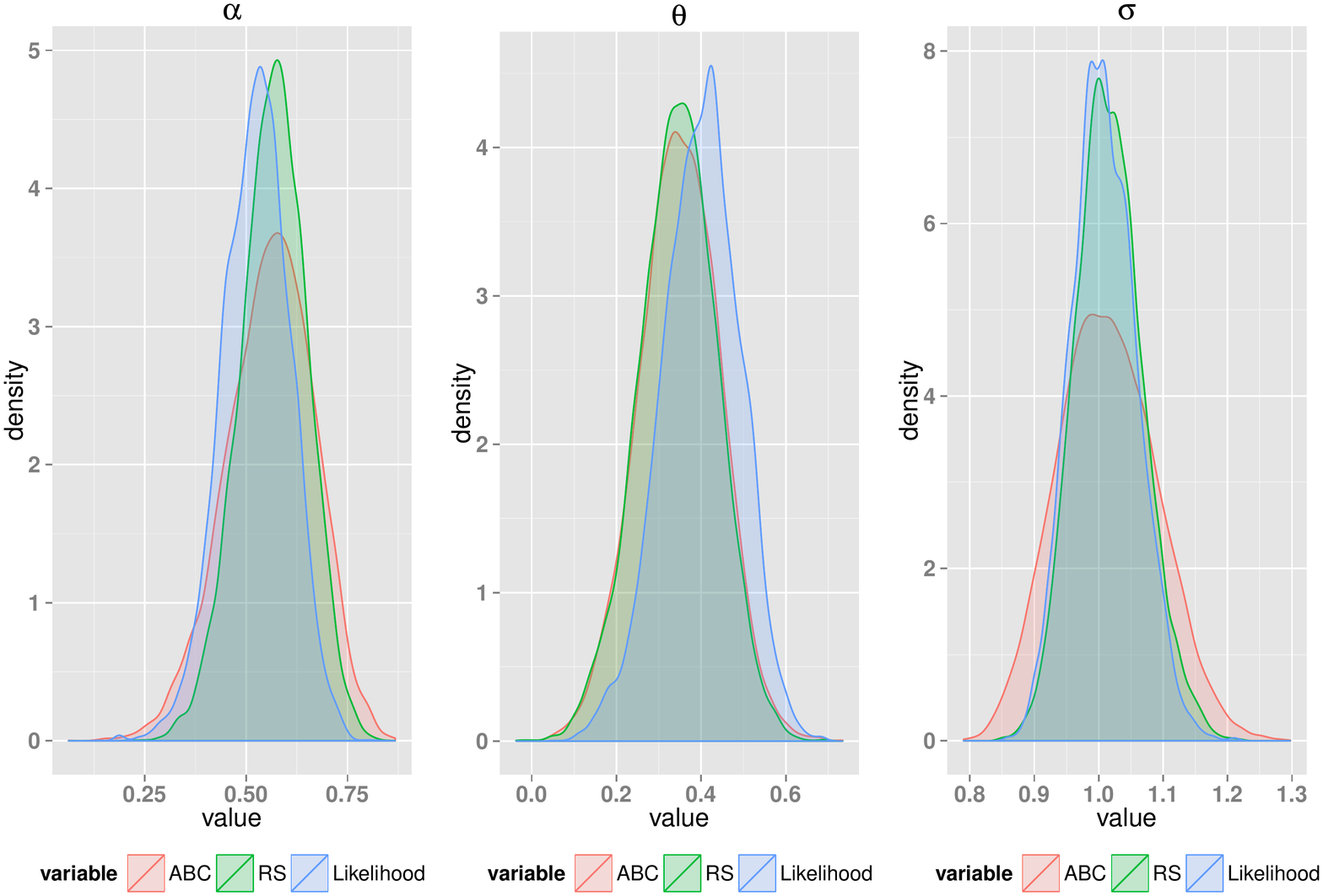}
\end{figure}

Figure \ref{fig:exponential} depicts a particular draw of the ABC posterior distribution (which coincides with the likelihood-based posterior since the statistics are sufficient), along with two  generated by the RS sampler. The first one uses the sample mean as auxiliary statistic and hence is exactly identified. The second uses two auxiliary statistics: the sample mean and the sample variance. For the AR-ABC, we draw from the prior ten million times and keep the ten thousand nearest draws. This  corresponds to a value of $\delta=0.0135$. For the RS, we draw
one million times\footnote{This means that we solve the optimization problem one million times. Given that the optimization problem is one dimensional, the one dimensional R optimization routine \textit{optimize} is used. It performs a combination of the golden section with parabolic interpolations. The optimum is found, up to a given tolerance level (the default is $10^{-4}$), over the interval $[0,10].$} and keep the ten thousand nearest draws which corresponds to a $\delta=0.0001.$  As for the weight matrix $W$, if we put $W_{11}>0$ and zero elsewhere, we will recover the exactly identified distribution. Here, we intentionally put a positive weight on the variance (which is  not  a sufficient statistic) to check the effect on the posterior mean. With $\mathrm W_{11}=1/5$ and $\mathrm W_{22}=4/5$, the RS posterior means
are 0.7452 and 0.7456 for the just and overidentifed cases. The corresponding values are
are 0.7456 and .7474 for the exact posterior and the ABC-AR. They are very similar.

\paragraph{Example 5: ARMA(1,1):}
For $t=1,\ldots, T=200$ and $\theta_0=(\alpha_0, \thetanormal_0,\sigma_0)= (0.5,0.5,1.0)$,
 the data are generated as
\[ y_t = \alpha y_{t-1} + \varepsilonnormal_t+\thetanormal\varepsilonnormal_{t-1}, \quad\quad
\varepsilon_t \sim \mathcal{N}(0,\sigma^2).\]
Least squares estimation of the auxiliary model
\[y_t = \phi_1 y_{t-1}+\phi_2 y_{t-2}+\phi_3 y_{t-3}+\phi_4 y_{t-4}+u_t\]
yields $L=5> K=3$ auxiliary parameters
\[\hat \psi = (\hat \phi_1,\hat \phi_2,\hat \phi_3,\hat \phi_4,\hat \sigma_u^2).\]
We let  $\pi(\alpha,\thetanormal,\sigma)=\mathbb{I}_{\alpha,\thetanormal \in [-1,1],\sigma \geq 0}$
and  $\mathrm{W}=\mathrm{I}_5$ which is inefficient. In this example,  $\hat\psi$ are not sufficient statistics
since $y_t$  has an infinite order autoregressive representation.

 We draw $\sigma$ from a uniform distribution on $[0,3]$ since $\mathcal{U}_
 {[0,\infty]}$ is not a proper density. The weights of the RS are obtained by numerical differentiation. The likelihood based  posterior is computed by MCMC using the Kalman Filter  with  initial condition $\varepsilonnormal_0=0.$   As mentioned above,  the desired number of draws is obtained by setting the quantile instead of setting the tolerance $\delta$.
For the RS, we keep the 1/10=10\% closest draws corresponding to a $\delta=0.0007$.
The Sequential Monte-Carlo implementation
of ABC (SMC-ABC)  is more efficient at targeting the posterior than the ABC-AR.
Hence we also compare the RS with SMC-ABC as implemented in the Easy-ABC package
of \citet{ljd:13}.\footnote{We implemented the SMC-ABC in two ways. First, we use the procedure
in\citet {vo-drovandi} using code generously provided
by Christopher Drovandi. We also use the Easy-ABC package in $R$ of \citet{ljd:13}.
We thank an anonymous referee for this suggestion.}  The requirement
for 10,000 posterior draws are as follows:
\begin{center}
\begin{tabular}{c|cccc} \hline
  & AR-ABC & SMC-ABC & RS & Likelihood\\ \hline
Computation Time (hours) & 63 & 25 & 5 & 0.1\\ \hline
Effective number of draws & 100,000,000 & 36,805,000 & 10,153,108 & \\ \hline
$\delta$ & 0.0132 & 0.0283 & 0.0007 &  \\\hline \hline
\end{tabular}
\end{center}
%required to compute the $10,000$ posterior draws is 5 hours for the RS, 25 hours for the SMC-ABC and 6 minutes using the likelihood. This large difference in time between the RS and the SMC-ABC can be partly traced back to the number of times we have to simulate the model: 10,153,108 for the RS and 36,805,000 for the SMC-ABC.
The difference, both in terms of computation time and number of model simulations,
is notable. As shown in figure \ref{fig:arma} the quality of the approximation
is also different, especially for $\alpha$ and $\sigma$. The difference can
be traced to $\delta$. The $\delta$ used for the SMC-ABC is effectively much
larger than for the RS. A better approximation requires a smaller $\delta$
which implies longer computational time. Alternatively stated,
the acceptance rate at a low value of $\delta$ is very low. The caveat is that
the speed gain is possible only if the optimization problem can
be solved in a few iterations and reasonably fast. In practice, there will
be a trade-off between the number of draws and the number of iterations in
the optimization step as we further explore below.

\section{Acceptance Rate}
 The RS was initially developed in \citet{jjng-14} as a framework to help understand frequentist (SMD) and the Bayesian (ABC) way of likelihood-free estimation. But it turns out that the RS has one computation advantage that is worth highlighting. The issue pertains to the low acceptance rate of the ABC.

 As noted above, the ABC exactly recovers the posterior distribution associated with the infeasible likelihood if $\hat\psi$ are sufficient statistics
and $\delta=0$ as noted in \citet{blum-10}.
Of course, $\delta=0$ is an event of measure zero, and the ABC  has an approximation
bias that depends on
  $\delta$. In theory, a small $\delta$ is desired. The ABC  needs a  large number of draws to  accurately approximate the posterior and can be computationally costly.

To illustrate this point, consider
 estimating the mean $m$ in Example 3 with $\sigma^2=1$ assumed to be
 known,
 and $\pi(m) \propto 1$.  All computations are based on the software package \textsc{R}.
 From a previous draw $m^b$,  a random walk step gives
$m^\star=m^b+\varepsilon, \varepsilon \sim \mathcal{N}(0,1)$.
For small $\delta$,
we can assume $m^\star|\hat{m} \sim \mathcal{N}(\hat{m},1/T)$.
From a simulated
sample of $T$ observations,  we get an  estimated mean $\hat{m}^\star \sim
\mathcal{N}(m^\star,1/T)$.
As is typical of MCMC  chains,
these draws are serially
correlated. To see that  the algorithm can be stuck for a long time if $m^*$ is
far from $\hat m$,
observe that  the event
$\hat{m}^\star \in [\hat{m}-\delta, \, \hat{m} + \delta]$ occurs
with probability
\[\mathbb{P} ( \hat{m}^\star \in [\hat{m}-\delta, \,
\hat{m} + \delta] )
%= \mathbb{P} \left( \hat{m}^\star \leq \hat{m} +
%\delta \right) - \mathbb{P} \left( \hat{m}^\star \leq \hat{m} - \delta \right)
= \Phi \left( \sqrt{T}(\hat{m}+\delta-m^\star) \right)-\Phi \left(
\sqrt{T}(\hat{m}-\delta-m^\star) \right) \approx 2\sqrt{T}\delta \phi \left(
\sqrt{T}(\hat{m}-m^\star) \right).\]
The acceptance probability  $\int_{m^*}
\mathbb{P} ( \hat{m}^\star \in [\hat{m}-\delta, \,
\hat{m} + \delta] ) dm^*$ is thus approximately linear in $\delta$.
To keep the number of accepted draws constant,
we need to increase the number of draws as we decrease $\delta$.

This result that the
  acceptance rate is linear in $\delta$
also applies in the general case.
 Assume that
 $\hat{\psi}^\star(\theta^\star) \sim \mathcal{N}(\psi(\theta^\star),\Sigma/T)$.
 We  keep the draw if $\| \hat{\psi}-\hat{\psi}^\star(\theta^\star)\| \leq
 \delta$. The probability of this event can be bounded above by $
 \sum_{j=1}^K \mathbb{P}
 \left( |\hat{\psi}_j-\hat{\psi}_j^\star(\theta^\star)| \leq \delta \right)$ i.e.:
 \begin{eqnarray*}
  \sum_{j=1}^K \Phi
 \left( \frac{\sqrt{T}}{\sigma_j} \left(\hat{\psi}_j+\delta-\psi_j(\theta^\star)
 \right) \right) - \Phi \left( \frac{\sqrt{T}}{\sigma_j} \left(\hat{\psi}_j-\delta-\psi_j(\theta^\star) \right) \right)
 \approx 2\sqrt{T}\delta \sum_{j=1}^K \frac{\phi}{\sigma_j} \left( \frac{\sqrt{T}}{\sigma_j}
  \left( \hat{\psi}_j-\psi_j(\theta^\star) \right) \right).
  \end{eqnarray*} The acceptance probability is still at best  linear
in $\delta$. In general  we need to increase
   the number of draws at least as much as $\delta$ declines to keep the number
   of accepted draws fixed.

%   Also note that when $\psi(\theta^\star)$ is far from $\hat{\psi}$, the acceptance
%   probability will tend to be very low and the MCMC chain will stay around that value.

\begin{table}[ht]
\centering
\caption{Acceptance Probability as a function of $\delta$}

\label{tab:acceptance}
\begin{tabular}{llllll} \hline
 $ \delta $ & 10 & 1 & 0.1 & 0.01 & 0.001 \\ \hline
$\mathbb{P}(\|\hat{\psi}-\hat{\psi}^b \|_W \leq \delta)$ &
0.72171 & 0.16876 & 0.00182 & 0.00002 & $<$0.00001
\\
\hline \hline
\end{tabular}
%\begin{tabular}{rr}
%  \hline \hline
% $\delta$ & $\mathbb{P}(\|\hat{\psi}-\hat{\psi}^b \|_W \leq \delta)$ \\
%  \hline
%10 & 0.72171 \\
%1 & 0.16876 \\
%0.1 & 0.00182 \\
%0.01 & 0.00002 \\
%0.001 & 0.00000 \\
%
 %  \hline \hline
%\end{tabular}\\

\end{table}

Table \ref{tab:acceptance} shows the acceptance rate for Example 3
 for $\theta_0=(m_0,\sigma_0^2)=(0,2)$,  $T=20$, and weighting matrix
  $W=\text{diag}(\hat{\sigma}^2,2\hat{\sigma}^4)/T,  \pi(m,\sigma^2) \propto \mathbbm I_{\sigma^2 \geq 0}$.
The results confirm that for small values of $\delta$, the acceptance rate
is approximately linear in $\delta$.
%This observation about  $\delta$ has practical implications for the ABC-MCMC.
Even though in theory, the targeted ABC posterior should be closer to the
true posterior when $\delta$ is small, this may not be true in practice because
of the poor properties of the MCMC chain. At least for this example,  the
MCMC chain with moderate value of $\delta$ provides a better approximation
to the true posterior density.

To overcome the low acceptance rate issue, \citet{beaumont-zhang-balding}
suggests to  use local regression techniques to approximate $\delta=0$ without
setting it equal to zero. The convergence rate is then non-parametric. \citet
{gao-hong} analyzes the estimator of \citet{creel-kristensen-il} and
finds that  to compensate for the large variance
associated with the kernel smoothing,
the number of simulations need to be larger than $T^{K/2}$
 to achieve $\sqrt{T}$ convergence,  where $K$ is the
number of regressors.
Other methods  that aim to increase the acceptance rate   include the  ABC-SMC  algorithm
of \citet{sisson-fan-tanaka, sisson-fan}, as well as the adaptive weighting variant
due to \citet{bonassi-west-15},  referred to below as
SMC-AW. These methods  build a sequence of proposals to more efficiently target
the posterior. The acceptance rate still declines rapidly with $\delta$, however.

The RS circumvents this problem because each $\theta^b$ is accepted by virtue
of being the solution of an optimization problem, and hence  $\hat\psi-\hat\psi^b
(\theta^b)$ is the smallest possible. In fact, in the exactly identified case, $\delta=J_1^b=0$.  Furthermore, the sequence of optimizers are independent, and the sampler cannot be stuck. We use two more examples to highlight this feature.

%This problem appears not to be specific to   the
% the indicator function/rectangular kernel formulation of the ABC.
%\citet{gao-hong} analyzes the estimator of \citet{creel-kristensen-il} (which is
%effectively a noisy ABC estimator). They suggests that the problem arises
%because to compensate for the large variance
%associated with the kernel,
%the number of simulations need to be larger than $T^{K/2}$
% to achieve $\sqrt{T}$ convergence,  where $K$ is the
%number of regressors. To overcome this problem,
%  \citet{beaumont-zhang-balding} suggests local regression adjustment, while \citet{bonassi-west-15} suggests
% Sequential Monte-Carlo with adaptive weighting (hereafter SMC-AW). The RS circumvents this problem because each $\theta^b$ is accepted by virtue of being the solution of an optimization problem, and hence  $\hat\psi-\hat\psi^b(\theta^b)$ is the smallest possible. In fact, in the exactly identified case, $\delta=J_1^b=0$.  Furthermore, the sequence of optimizers are independent, and the sampler cannot be stuck. We use one more example to highlight this feature.

\begin{figure}[ht]

  \centering
  \caption{Mixture Distribution}
  \label{fig:mixture}
  \includegraphics[width=6.0in,height=3.5in]{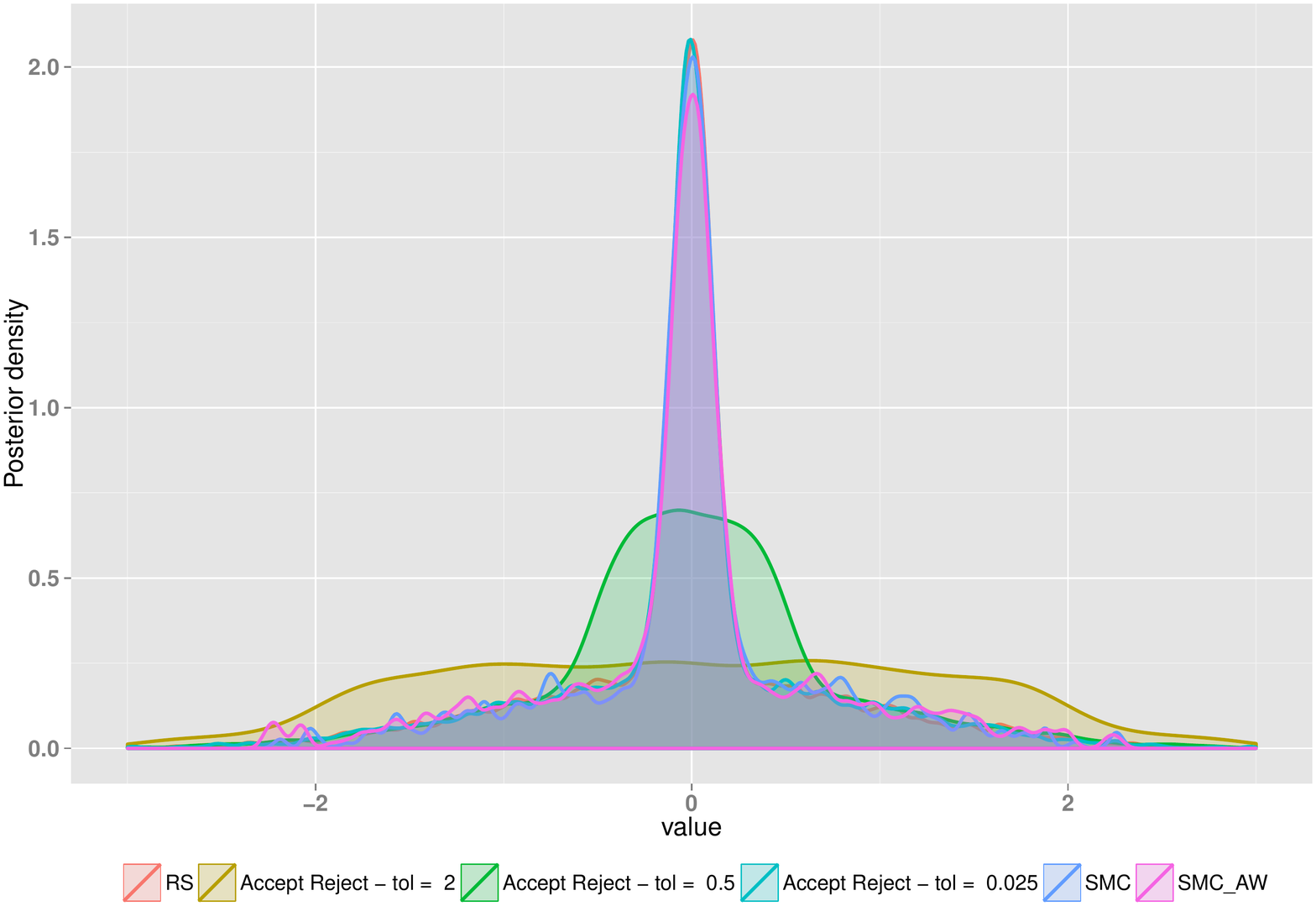}
\end{figure}

\paragraph{Example 6: Mixture Distribution}
 Consider the example in \citet{sisson-fan-tanaka}, also considered in \citet{bonassi-west-15}.
 Let    $\pi(\thetanormal) \propto 1_{\thetanormal \in [-10,10]}$ and
 \[x|\thetanormal \sim 1/2 \mathcal{N}(\thetanormal,1)+1/2 \mathcal{N}(\thetanormal,1/100)\]
 Suppose we observe one draw $x=0$. Then the true posterior is $\thetanormal|x \sim
 1/2 \mathcal{N}(0,1)+1/2 \mathcal{N}(0,1/100)$ truncated to $[-10,10]$. As
 in  \citet{sisson-fan-tanaka} and \citet{bonassi-west-15},
  we choose three tolerance levels: $(2,0.5,0.025)$ for AR-ABC.  Figure \ref{fig:mixture} shows that the
  ABC posterior distributions  computed using accept-reject sampling with $\delta=0.025$ are similar to the ones using SMC with and without adaptive weighting. The RS posterior distribution is  close to both ABC-SMC and ABC-SMC-AW, and all  similar to Figure 3 reported in \citet{bonassi-west-15}.
However, they are quite  different from the AR-ABC with $\delta=2$ and 0.5 are 2, showing that the choice of $\delta$ is important in ABC.

While the SMC,  RS, and ABC-AR sampling schemes can produce similar posterior distributions, Table \ref{tab:Time} shows that their computational time differ dramatically.  The two SMC algorithms need to sample from a multinomial distribution which are evidently more time consuming.    When $\delta=0.25$, the AR-ABC posterior distribution is close to the ones produced by the SMC samplers and the RS, but the computational cost is still high. The AR-ABC is computationally efficient when $\delta$ is large, but as seen from Figure \ref{fig:mixture}, the posterior distribution is quite poorly approximated. The RS takes 0.0017 seconds to solve, which is amazingly fast because  for this example, the solution is available analytically. No optimization is involved, and there is no need to evaluate the Jacobian because the model is linear. Of course, in cases when the SMD problem is numerically challenging, numerical optimization can be time consuming as well. Our results nonetheless suggest a role for optimization in  Bayesian computation; they need not be mutually exclusive. Combining the ideas is an interesting topic for future research.

\begin{table}[ht]
\centering
\caption{Computation Time (in seconds)} \label{tab:Time}

\begin{tabular}{r|rrr|rrr}
 RS &\multicolumn{3}{c|}{ABC-AR} & \multicolumn{2}{c}{ABC-SMC}\\ \hline
  &$\delta$=2 &  $\delta$=.5 &  $\delta$=.025 & Sisson et-al & Bonassi-West\\
\hline
.0017 & 0.4973 & 1.6353 & 33.8136 & 190.1510 & 199.1510\\
\hline
\end{tabular}
\end{table}

\paragraph{Example 7: Precautionary Savings}
The foregoing examples are simple and are serve illustrative purposes.  We
now consider an example that indeed has an infeasible likelihood.
In \citet{deaton-91}, agents maximize
expected utility $\mathbb{E}_0 \left(\sum_{t=0}^\infty \beta^t u(c_t) \right)$ subject to the constraint
that assets $a_{t+1}=(1+r)(a_t+y_t-c_t)$ are bounded
below by zero, where $r$ is interest rate, $y$ is income and $c$ consumption. The desire for precautionary saving interacts with borrowing
constraints to generate a policy function that is not everywhere concave, but
is a piecewise linear when cash-on-hand is below an endogenous threshold. The
 policy function can only be solved numerically at assumed parameter values.
SMD estimation thus consists of solving the model and simulating $S$ auxiliary
statistics at each guess
$\theta$. \citet{michaelides-ng} evaluate the finite sample
properties of several SMD estimators using a model with similar features.
Since the likelihood for this model is  not available analytically. Hence
Bayesian estimation of this model has not been implemented.
Here,
we use the RS to approximate the posterior distribution.

We generate $T=400$ observations assuming that $U(c)=\frac{c^{1-\gamma}-1}
{1-\gamma}$, $y_t\sim \text{ iid } \mathcal{N}(\mu,\sigma^2)$ with
$r=0.05$,  $\beta=10/11$, $\mu=100,\sigma=10, \gamma=2$ as true values.
We estimate  $\theta=(\gamma,\mu,\sigma)$ and assume $(\beta,r)$ are known.
We use 10 auxiliary statistics:
\begin{align*}
  \hat \psi = \begin{pmatrix}
  \bar y &
  \hat\Gamma_{yy}(0) &\hat\Gamma_{aa}(0) &\hat\Gamma_{cc}(0) &
  \hat\Gamma_{cc}(1) &\hat\Gamma_{aa}(1) &
  \hat\Gamma_{cc}(2) &\hat\Gamma_{aa}(2) &
  \hat\Gamma_{cy}(0) &\hat\Gamma_{ay}(0)
%  \frac{1}{T}\sum_t (x_t-\bar x) (x_{t-j}-\bar x)\\
%  \frac{1}{T} \sum_{t} (y_t-\bar y)^2\\
%  \frac{1}{T} \sum_{t} (a_t-\bar a)^2\\
%  \frac{1}{T} \sum_{t} (c_t-\bar c)^2\\
%  \frac{1}{T} \sum_{t} (c_t-\bar c)(c_{t-1}-\bar c)\\
%  \frac{1}{T} \sum_{t} (a_t-\bar a)(a_{t-1}-\bar a)\\
%  \frac{1}{T} \sum_{t} (c_t-\bar c)(c_{t-2}-\bar c)\\
%  \frac{1}{T} \sum_{t} (a_t-\bar a)(a_{t-2}-\bar a)\\
%  \frac{1}{T} \sum_{t} (c_t-\bar c)(y_{t}-\bar y)\\
 % \frac{1}{T} \sum_{t} (a_t-\bar a)(y_{t}-\bar y)\\
  \end{pmatrix}^\prime
\end{align*}
where $\hat \Gamma_{ab}(j)=\frac{1}{T}\sum_
{t=1}^T (a_t-\bar a)(b_{t-j}-\bar b)$.
We generate $B=13,423$ draws  and keep the $3,356$ (25\%) nearest draws
to $\hat \psi.$ After weighting using the volume of the Jacobian matrix we have
an effective sample size of $1,421$ draws.\footnote{The effective sample size
is computed as $1/\sum_{b=1}^B w_b^2$ where the weights satisfy $\sum_{b=1}^B w^b =1.$}
We use an identity weighting matrix  so $J_{RS}(\theta)=\bar g(\theta)^\prime
\bar g(\theta)$.  The Jacobian is computed using finite differences for the RS.
As benchmark, we also compute an  SMD with $S=100$, $J_S=\bar g_S(\theta)^\prime
 \bar g_S(\theta)$.
 % using both an identity matrix and the inverse of the long-run variance of $\bar
%g(\theta)$ as $W$.  The latter is estimated
%using the method of Newey-West with a  lag length of $[4 (T/100)^{2/9}]=5$.
%In this exercise, the SMD has  two advantages over the RS.
%The first is  a more efficient
%weighting matrix.
In this exercise,  the SMD  only needs to solve
for the policy function once at each step of the optimization. Hence
the binding function can be approximated using simulated data at a low cost.
%A solution would be to solve the policy function on a grid of points and interpolate as in Deaton and Laroque (???), instead of solving the policy function at each $\theta$.
For this example, the programs are coded in \textsc{python}. The Nelder-Mead method is used for optimization.

\begin{table}[ht]
  \caption{Deaton Model: RS, SMD with $\mathrm{W}=\mathrm{I}$}
  \label{tab:Deaton}
  \centering
  \begin{tabular}{c|ccc|ccc}
     &  \multicolumn{3}{c|}{Posterior Mean/Estimate} & \multicolumn{3}{c}
     {Posterior SD/SE} \\ \hline
     & $\gamma$ & $\mu$ & $\sigma$ & $\gamma$ & $\mu$ & $\sigma$ \\ \hline
     RS            & 1.86 & 99.92 & 10.48 & 0.19 & 0.84 & 0.37 \\ \hline
     SMD   & 1.76 & 99.38 & 10.31 & 0.12 & 0.60 & 0.34 \\ \hline
%     SMD Optimal   & 1.89 & 95.81 & 8.90  & 0.15 & 0.28 & 0.21 \\ \hline
     \hline
  \end{tabular}
\end{table}

\begin{figure}[ht]
  \caption{Deaton Model: RS and SMD}
  \label{fig:DEATON}
  \includegraphics[width=6in, height=3.50in]{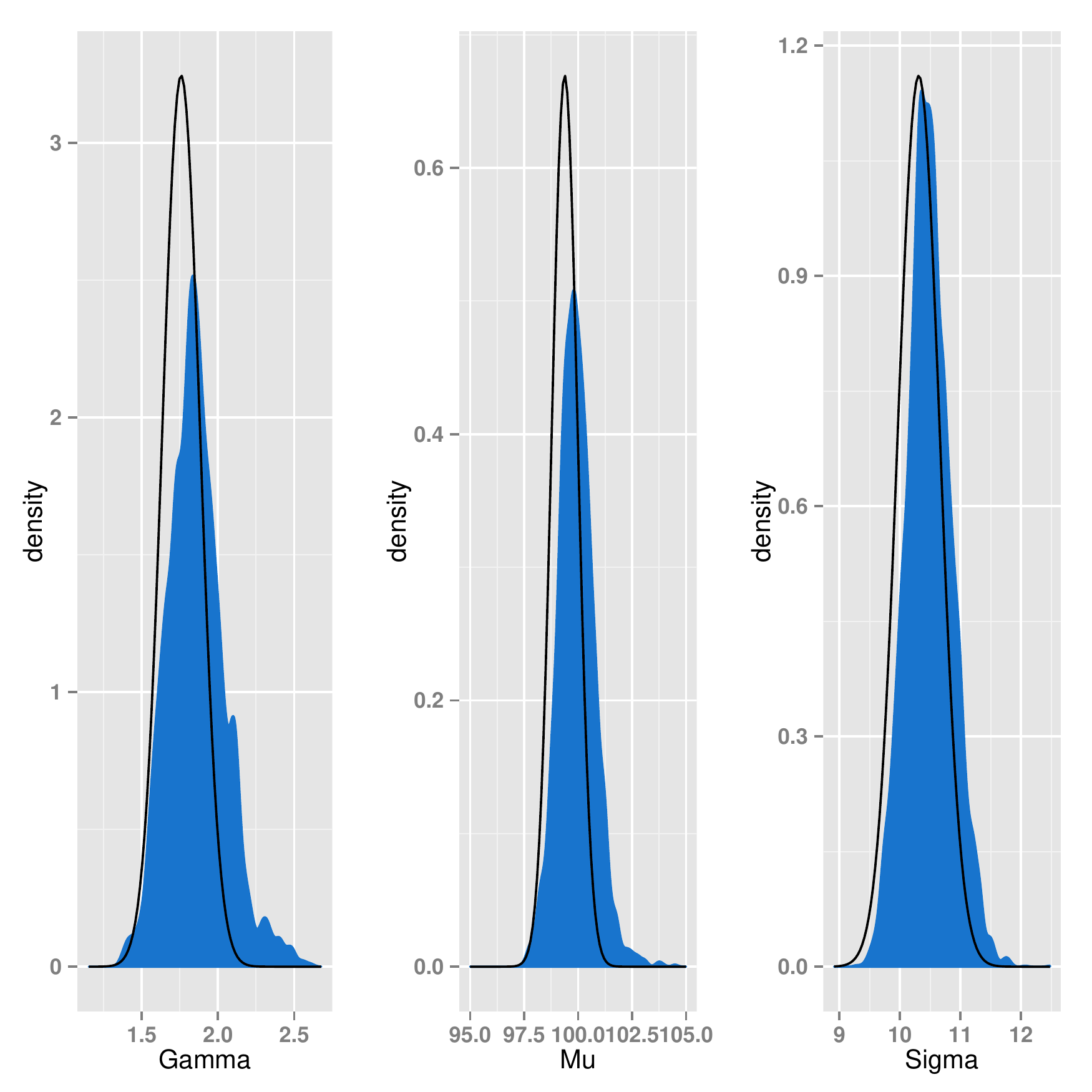}\\
  \textit{Note:} Blue density: RS posterior, Black line: large sample approximation for the SMD estimator (identity weighting matrix).
\end{figure}

Figure \ref{fig:DEATON} shows the posterior distribution of the RS (blue)
along with the SMD distribution (purple) as approximated by $\mathcal{N}(\hat \theta_\text
{SMD},\hat{\mathrm{V}}_\text{SMD}/T)$ according to asymptotic theory.
Table \ref{tab:Deaton} shows that the two sets of point estimates are similar.
As explained in \citet{jjng-14}, the SMD uses simulations to approximate the
binding function
while the RS (and by implication the ABC) uses simulations to approximate
the infeasible posterior distribution. In this example, the
difference in bias is quite small. We should
note that the RS took well over a day to solve while the SMD took less than three hours to compute. Whether we use our own code
for the ABC-MCMC or from packages available, the acceptance rate is too low
for the exercise to be feasible.
%To put this result into perspective, note that
%since $y$ is exogenous and does not depend on $c$ or $a$, it could also be
%estimated by MLE using data on $y$ alone.
%Since $y_t \sim \mathcal{N}(100,100)$,  we can view $\bar y$ as the MLE for
%$\mu$, and
% $\frac{\sigma}{\sqrt{T}} = 0.5$  as an efficiency bound.

\section{Conclusion}
This paper studies properties of the reverse sampler considered in \citet
{jjng-14} for  likelihood-free estimation. The sampler produce draws from  the infeasible  posterior distribution by solving  a sequence of frequentist SMD problems. We showed that the reverse sampler uses the Jacobian matrix as importance ratio.  In  the over-identified case, the importance ratio can be computed using the volume of the Jacobian matrix. The reverse sampler does not suffer from the problem of low acceptance rate that makes the ABC computationally demanding.

\clearpage
\baselineskip=12.0pt
\bibliography{metrics,kn,macro,consum}

\clearpage

\clearpage

\FloatBarrier

\end{document}